\documentclass[11pt, a4paper]{article}
\usepackage[english]{babel}
\usepackage[utf8]{inputenc}
\usepackage[T1]{fontenc}
\usepackage[a4paper, margin=0.7in]{geometry}
\usepackage{blindtext}
\usepackage{graphicx}
\usepackage{wrapfig}
\usepackage{hyperref}
\usepackage{datetime2}
\usepackage[table]{xcolor}
\usepackage{multirow}
\usepackage{amsmath}
\usepackage{amsthm}
\usepackage{algorithm}
\usepackage{algpseudocode}
\usepackage{lipsum}
\usepackage[shortlabels]{enumitem}
\usepackage[toc,
  nonumberlist
]{glossaries}
\usepackage{caption}
\usepackage{subcaption}
\usepackage[makeroom]{cancel}

\hypersetup{
    colorlinks=true,
    linkcolor=blue,
    filecolor=magenta,      
    urlcolor=blue,
    pdftitle={Proxima},
}

\graphicspath{ {./images/} }
\newtheorem{statement}{Statement}[section]
\newtheorem{definition}{Definition}[section]
\newtheorem{conclusion}{Conclusion}[section]
\newtheorem{observation}{Observation}[section]
\newtheorem{assumption}{Assumption}[section]

\title{Proxima: a DAG--based cooperative distributed ledger}

\author{
  Evaldas Dr\k{a}sutis\footnote{Contacts: email: evaldas.drasutis@mif.vu.lt, GitHub: lunfardo314, X/Twitter: @lunfardo314}\\
}

\date{\today}

\makeglossaries
\newglossaryentry{deterministic}{
    name=deterministic,
    description={is a trait in the distributed system, which assumes that every participant perceives (interprets) a particular data structure or other things exactly the same way}
}    
\newglossaryentry{non-deterministic}{
    name=non-deterministic,
    description=is the opposite to being \gls{deterministic}. i.e. perception of a data structure depends on the participant and the random factors
}
\newglossaryentry{pow}{
    name=PoW,
    description={proof-of-work, a principle of \gls{probabilistic consensus} introduced by Satoshi Nakamoto in Bitcoin}
}
\newglossaryentry{pos}{
    name=PoS,
    description={proof-of-stake, a consensus principle used in distributed crypto ledgers. By our definition, it includes dPoS (delegated PoS) and other variations. We assume PoS requires Byzantine agreement on the stakes of participants, therefore inherently rely on \gls{bft} consensus protocols (admittedly, there may be broader definitions of PoS)}
}
\newglossaryentry{UTXO ledger}{
    name=UTXO ledger,
    description={is a name of the model of the crypto ledger, where ledger state consists of \glspl{UTXO}. It is an alternative to the \textbf{account-based} ledger model, where ledger state consists of accounts}
}
\newglossaryentry{UTXO}{
    name=UTXO,
    description={stands for \textbf{U}nspent \textbf{T}ransa\textbf{X}tion \textbf{O}utput, an one-time non-fungible asset on the \gls{ledger state}, controlled by some \gls{token holder}. In many contexts it is a synonym of \gls{output}. See \ref{sec:utxotx}}
}
\newglossaryentry{output}{
    name=output,
    description={in the context of the \gls{UTXO ledger} model, it is a synonym of \gls{UTXO}}
}

\newglossaryentry{bft}{
    name=BFT,
    description={\textit{Byzantine Fault Tolerant} consensus protocol, which stems from seminal works of Lamport et al.\footnote{The Byzantine Generals Problem \url{https://dl.acm.org/doi/10.1145/357172.357176}}. We assume BFT is inherently based on globally known set of participants, also known as \textit{committee}. The \gls{pos} systems  at their core are usually built around BFT consensus protocols}
}
\newglossaryentry{DAG}{
    name=DAG,
    description={directed acyclic graph, a mathematical model for partially ordered structures. In the context of \glspl{crypto ledger}, it is used as an alternative to the totally ordered structure of the blockchain}
}
\newglossaryentry{probabilistic consensus}{
    name=probabilistic consensus,
    description={is a consensus principle which makes a set of participants converge to the same perception of the \gls{ledger state}  with big probability but without guarantee of reaching it. It is different from the \gls{bft} consensus due to assumed unbounded and \gls{non-deterministic} set of participants in the consensus}
}
\newglossaryentry{transaction}{
    name=transaction,
    description={in the context of Proxima, it is a UTXO transaction: a \gls{deterministic} prescription which transforms a \gls{ledger state} by removing particular \glspl{UTXO} (consumed outputs) from it and creating new \glspl{UTXO} (produced outputs) on the next one}
}
\newglossaryentry{cooperative consensus}{
    name=cooperative consensus,
    description={is a principle of \gls{probabilistic consensus} introduced in this paper. It preserves \gls{permissionless decentralization} without employing \gls{pow}}
}
\newglossaryentry{permissionless decentralization}{
    name=permissionless decentralization,
    description={is a trait of a distributed \gls{crypto ledger} where consensus on the \gls{ledger state} is achieved by a permissionless, unbounded and generally unknown set of participants. We see this trait as the defining trait of the \gls{Nakamoto consensus}. \gls{pos} crypto ledgers usually does not have this trait}
}
\newglossaryentry{Nakamoto consensus}{
    name=Nakamoto consensus,
    description={is usually used as a synonym of \gls{probabilistic consensus}. The Nakamoto consensus has the trait of \gls{permissionless decentralization}. Proxima introduces the \gls{cooperative consensus}, a \gls{Nakamoto consensus} which does not use \gls{pow}}
}
\newglossaryentry{blockchain}{
    name=blockchain,
    description={is a data structure, which represents a tree of \textit{blocks}, whereas each block represents a \gls{deterministic} update to the ledger. Blockchain always have single \textit{longest chain} of blocks which represents the most probable consensus ledger state. Blocks which does not belong to the longest chain are ultimately \gls{orphaned}, i.e. are not included into the history of the ledger. The \gls{crypto ledger} and \gls{DLT} are commonly used as synonyms of \textit{blockchain} in most contexts}
}
\newglossaryentry{crypto ledger}{
    name=crypto ledger,
    description={in the context of this paper, a synonym of \gls{DLT}}
}
\newglossaryentry{DLT}{
    name=DLT,
    description={stands for the \textit{Distributed Ledger Technology}, a generalization of terms \gls{blockchain} and \gls{crypto ledger}}
}

\newglossaryentry{ledger coverage}{
    name=ledger coverage,
    description={is a \gls{deterministic} value, calculated for each \gls{sequencer transaction}. It is a sum of amounts of \glspl{UTXO} consumed by the \gls{past cone} of the sequencer transaction in its \gls{baseline state}. See more in  \ref{def:coverage}}
}
\newglossaryentry{biggest ledger coverage}{
    name=biggest ledger coverage,
    description={is a rule followed by \glspl{sequencer} in the \gls{cooperative consensus}. Sequencers produce \glspl{sequencer transaction} by maximizing their \gls{ledger coverage}. In Proxima, \textit{biggest ledger coverage} rule plays same role as the \textit{longest chain rule} in the \gls{pow} consensus}
}
\newglossaryentry{token holder}{
    name=token holder,
    description={is an entity which controls assets on the \gls{ledger} with their private keys}
}
\newglossaryentry{sequencer}{
    name=sequencer,
    description={is a particular strategy of the \gls{token holder} on the Proxima ledger. It allows the token holder to gain inflation and other rewards by pro-actively consolidating other \gls{UTXO} transactions into the \gls{ledger state} through cooperation with other token holders. Technically, the \textit{sequencer} is a program, which have access to the private key of the \gls{token holder} and builds \gls{chain} of \glspl{sequencer milestone} on behalf of the token holder. See Section \ref{sec:sequencing}.\nameref{sec:sequencing}}
}
\newglossaryentry{UTXO tangle}{
    name=UTXO tangle,
    description={is a \gls{transaction DAG} with output consumption and \gls{endorsement} relations as its edges. It is the main data structure of the \gls{cooperative consensus} and an alternative to \gls{blockchain}. \Gls{past cone} of every vertex of the \textit{UTXO tangle} represents a consistent \gls{UTXO ledger}}    
}
\newglossaryentry{transaction DAG}{
    name=transaction DAG,
    description={is a \gls{DAG} with \glspl{transaction} as vertices. In this paper, it is \gls{UTXO tangle}. In contrast to the \textit{block DAG}, the \textit{transaction DAG} is built by \glspl{token holder} rather than \textit{block producers}}    
}
\newglossaryentry{liquidity}{
    name=liquidity,
    description={is the ability to convert fungible crypto tokens native to the ledger to fiat currency and/or goods any time and in a permissionless way. Liquidity of the crypto token is equivalent to saying that the token has real world value}    
}
\newglossaryentry{endorsement}{
    name=endorsement,
    description={is an asymmetric relation (a directed edge in the \gls{DAG}) between two \gls{sequencer} transactions. In addition to the consumption relation between UTXO transactions, endorsements form edge set of the \gls{DAG}. Endorsements in the Proxima ledger model is en extension of the well known UTXO ledger models such as the ones used by of Bitcoin and Cardano (EUTXO). Technically, \textit{endorsement} is a \textit{hash} of the parent transaction and is part of the signed data of the transaction}
}
\newglossaryentry{past cone}{
    name=past cone,
    description={is a sub-graph of the \gls{UTXO tangle}, reachable from a certain vertex. Usually it is considered with respect to a chosen \gls{baseline state}. In the latter case, all transactions which belong to the baseline state, are considered \textit{rooted} and are excluded from the past cone of consideration. Past cone is \gls{deterministic}, i.e. when fully reconstructed (solidified), it is perceived equally by all participants in the asynchronous distributed system}    
}
\newglossaryentry{baseline state}{
    name=baseline state,
    description={also, \textit{baseline ledger state}. A chosen \gls{ledger state}, from which the \gls{past cone} of certain transaction $T$ can be built by incrementally applying transactions. \textit{Baseline state} is usually specified as a \gls{branch transaction}. For a given $T$, more than one possible baseline states exist and they form a chain of branches. \Gls{genesis ledger state} can be a baseline for any $T$ in the \gls{UTXO tangle}. In the context of \gls{sequencer milestone} $T$, the baseline state is assumed to be a \gls{branch transaction} with the latest \gls{timestamp} in the \gls{past cone} of $T$}
}
\newglossaryentry{ledger state}{
    name=ledger state,
    description={in general, is a list of accounts or non-fungible assets with their corresponding owners. Ledger state of the \gls{UTXO ledger} is a set of \glspl{UTXO} (also known as \textbf{UTXO set}). \Gls{past cone} of every vertex of the \gls{UTXO tangle} represents a \gls{ledger} \textbf{and} a ledger state}
}
\newglossaryentry{genesis ledger state}{
    name=genesis ledger state,
    description={a list of \glspl{UTXO}, from which the \gls{UTXO tangle} and, therefore, any other \gls{ledger state} in it, can be build by incrementally adding \glspl{transaction}}
}
\newglossaryentry{ledger}{
    name=ledger,
    description={in general, is a set of transactions, which transforms \gls{genesis ledger state} to a particular \gls{ledger state}. \Gls{UTXO ledger} consists of UTXO \glspl{transaction}}
}
\newglossaryentry{constraint}{
    name=validity constraint,
    description={or just \textbf{constraint} is a validation script (formula) in \textit{EasyFL} language, an immutable part of the \gls{UTXO}. It is evaluated in the context of the transaction, which consumes or produces it. A \gls{transaction} is valid if and only if all constraints in its context evaluates to \textit{true}}
}
\newglossaryentry{chain constraint}{
    name=chain constraint,
    description={is a specific \gls{constraint} script attached to the \gls{UTXO}. It is tagged with unique \textit{ID}, which becomes the immutable \textit{ID} of the \gls{chain}. \textit{Chain constraint} enforces every consuming transaction to produce exactly one successor: a \gls{chained output} with the same \textit{ID}}
}
\newglossaryentry{chained output}{
    name=chained output,
    description={is a \gls{UTXO} with \gls{chain constraint} attached to it. \textit{Chained output} represents a persistent non-fungible asset on the ledger, uniquely identified by the \textit{chain ID}}
}
\newglossaryentry{chained transaction}{
    name=chained transaction,
    description={is a \gls{transaction} which produces or consumes a \gls{chained output}}
}
\newglossaryentry{chain}{
    name=chain,
    description={in the context of the \gls{UTXO tangle}, a sequence of \glspl{chained transaction}. Chains cannot be ``forked`` in the ledger due to the \glspl{chain constraint}}
}
\newglossaryentry{sequencer milestone}{
    name=sequencer milestone,
    description={is a \gls{chained output} with additional \textit{sequencer constraints} on it (see in the text). \textit{Sequencer milestones} form \textit{sequencer chain}. \textit{Sequencer milestones} are issued by \glspl{sequencer}}
}
\newglossaryentry{sequencer transaction}{
    name=sequencer transaction,
    description={a \gls{transaction} which produces a \gls{sequencer milestone}}
}
\newglossaryentry{branch transaction}{
    name=branch transaction,
    description={also \textit{branch}. A \gls{sequencer transaction} on a \gls{slot} edge. \textit{Branch transaction} represents a \gls{ledger state} committed to the database of the node in the form of the \textit{Merkle root} of the \gls{ledger state}. Branch transactions are used as \glspl{baseline state} of \glspl{sequencer transaction}. Two branch transactions, when one is not a descendant of another, are \glspl{conflict} by design, i.e. consumes same \textit{output} in their \glspl{past cone}}
}
\newglossaryentry{stem}{
    name=stem output,
    description={or \textit{stem} is a \gls{UTXO} with special \textit{stem constraint} on it. Each \gls{branch transaction} consumes exactly one stem output (predecessor) and produces exactly one successor of it. This makes all branch transactions conflicting. The purpose of the \textit{stem} is to enforce tree structure among branches}
}
\newglossaryentry{double-spend}{
    name=double-spend,
    description={also \gls{conflict}. It is a situation when several \glspl{transaction} consume same \gls{UTXO}. \textit{Double-spend} can't happen in a consistent \gls{ledger}. The \gls{ledger} is always \textit{conflict-free}, however the \gls{UTXO tangle} commonly contains multiple \textit{conflicts}. Each \textit{double-spend} means ``fork``, a split of the ledger to several conflicting ledgers}
}
\newglossaryentry{conflict}{
    name=conflict,
    description={see \gls{double-spend}}
}
\newglossaryentry{tick}{
    name=tick,
    description={is the smallest value on the \gls{ledger time} axis}
}
\newglossaryentry{slot}{
    name=slot,
    description={is 128 \glspl{tick} on the \gls{ledger time} axis}
}
\newglossaryentry{ledger time}{
    name=ledger time,
    description={is a logical clock, similar to \textit{block height}. Smallest value of the \textit{ledger time} axis is one \gls{tick}}
}
\newglossaryentry{timestamp}{
    name=timestamp,
    description={is a value of the \gls{ledger time} stored as a part of the \gls{transaction}. \textit{Timestamp} is $0$ \glspl{tick} at genesis. It is said timestamp $\tau$ \textit{is on the \gls{slot} edge} when $\tau \mod 128 = 0$}
}
\newglossaryentry{orphaned}{
    name=orphaned,
    description={means ultimately not included into the \gls{ledger} of consideration. In a \gls{probabilistic consensus}, being orphaned is a \gls{non-deterministic} trait}
}
\newglossaryentry{tag-along}{
    name=tag-along,
    description={means sending small amount, a \gls{tag-along fee} to a \gls{sequencer} of choice to motivate it to include the transaction into its \gls{ledger}. The \gls{tag-along fee} is sent as an amount in the \gls{tag-along output}}
}
\newglossaryentry{tag-along fee}{
    name=tag-along fee,
    description={small amount on the \gls{UTXO}, a fee ("bribe"), which motivates \gls{sequencer} to consume the output}
}
\newglossaryentry{tag-along output}{
    name=tag-along output,
    description={is an \gls{UTXO} with the amount of \gls{tag-along fee} sent to a \gls{sequencer}}
}
\newglossaryentry{delegation}{
    name=delegation,
    description={Making tokens available to the \gls{sequencer} so that it could include them into the \gls{ledger coverage} of their transactions but without possibility to steal them}
}
\newglossaryentry{inflation}{
    name=inflation,
    description={creating tokens on the \gls{ledger} "out of thin air", as enabled by the mechanism of \glspl{chain}. Inflation increases total supply of tokens on the \gls{ledger state}. See Section \ref{sec:inflation}}
}
\newglossaryentry{branch inflation bonus}{
    name=branch inflation bonus,
    description=is an {enforced verifiably random value of \gls{inflation} on the \gls{branch transaction}}
}

\begin{document}
\maketitle

\begin{abstract}
This paper introduces a novel architecture for a distributed ledger, commonly referred to as a \gls{blockchain}, which is organized in the form of a directed acyclic graph with \gls{UTXO} transactions as vertices, rather than as a chain of blocks. The consensus on the state of ledger assets is achieved through \gls{cooperative consensus}: an emerging profit-driven behavior of token holders themselves, which is viable only when they cooperate by following the \gls{biggest ledger coverage} rule. The cooperative behavior is facilitated by enforcing purpose-designed UTXO transaction validity constraints.

Token holders are the only category of participants authorized to make amendments to the \gls{ledger}, making participation completely permissionless - without miners, validators, committees, or stakes - and without any need of knowledge about the composition of the set of all participants in the consensus.

The setup allows to achieve high throughput and scalability together with low transaction costs, while preserving key aspects of \gls{permissionless decentralization} and asynchronicity found in Bitcoin and other proof-of-work blockchains, but without huge energy consumption. Sybil protection is achieved similarly to proof-of-stake blockchains, using tokens native to the ledger, yet the architecture operates in a leaderless manner without block proposers and committee selection.\footnote{
I am grateful to Serguei Popov, Bartosz Kuśmierz and @buddhini for constructive criticism, which I believe made the whole thing better.
}
\end{abstract}
\hfill \break \break \break \break \break
\begin{center}
``Large numbers of strangers can cooperate successfully by believing in common myths``\footnote{Yuval Noah Harari, Sapiens: A Brief History of Humankind}    
\end{center}

\begin{center}
``Blockchains are a tool for coordinating social consensus``\footnote{X/Twitter post by Dankrad Feist, an Ethereum researcher}
\end{center}

\clearpage
\tableofcontents

\clearpage

% ============================================================================
\section{Preface}

This whitepaper presents the Proxima design rationale at the "detailed technical vision" level. It is not intended to be an academic paper or a specification. Here, we follow a more engineering approach. To support design decisions presented in the whitepaper, we developed a testnet version of the Proxima node\footnote{Proxima repository: \url{https://github.com/lunfardo314/proxima}}. This whitepaper is partially based on findings while working on the node prototype and more advanced versions.

The ideas presented in the whitepaper are inspired by two main precursors: the famous \textit{Bitcoin whitepaper} by Satoshi Nakamoto\footnote{Bitcoin whitepaper: \url{https://bitcoin.org/bitcoin.pdf}} and the ideas presented in \textit{``The Tangle``} whitepaper by Serguei Popov\footnote{The Tangle: \url{https://assets.ctfassets.net/r1dr6vzfxhev/2t4uxvsIqk0EUau6g2sw0g/45eae33637ca92f85dd9f4a3a218e1ec/iota1_4_3.pdf}}. We adhere to the fundamental principles of Bitcoin design, especially the asynchronous, unbounded, and permissionless nature of the \gls{pow} consensus. The \gls{transaction DAG} consensus principle, "each transaction confirms other two", is taken from \textit{The Tangle} paper. We extend the original ideas of the \textit{tangle} with concepts of \gls{deterministic} \gls{UTXO ledger}, which are naturally based on \gls{DAG}. We pivot the entire reasoning around the DAG-based consensus from probabilistic to behavioral by introducing on-ledger incentives and purposefully designed ledger validity constraints, in the manner similar to Bitcoin and other blockchains. The proposed principle is called \gls{cooperative consensus}.

With Proxima, we make no claims about solving all blockchain problems such as trilema or being better than any other solutions. However, we hope that it will attract interest from blockchain system developers and academia. The \textit{ cooperative consensus} is one of just a few alternatives to the \gls{pow} consensus principle, which preserves the trait of \gls{permissionless decentralization}, a \textit{unique selling point} of \glspl{crypto ledger} in general. Some aspects, convergence and security in particular, certainly require deeper mathematical modeling and analysis. \\ 

Disclaimer. Proxima is not an IOTA fork. After 2024, when IOTA pivoted and abandoned its efforts known as \textit{coordicide}\footnote{The Coordicide \url{https://files.iota.org/papers/20200120_Coordicide_WP.pdf}} and \textit{IOTA 2.0}\footnote{IOTA 2.0. All you need to know \url{https://blog.iota.org/iota-2-0-all-you-need-to-know/}}, Proxima independently continues development of the advanced DAG-based crypto ledger, started by IOTA, by approaching the problem from scratch, i.e. from the first principles of the tangle. 

\section{Introduction}
\label{sec:introduction}

\subsection{Generic principles of the permissionless distributed ledger}
\label{sec:generic}

The \gls{crypto ledger} with the trait of \gls{permissionless decentralization} can be seen as a for-profit game among an unbounded set of participants with generally unrestricted behavior. The outstanding feature is the existence of a simple rule, an optimal strategy, which leads to the game-theoretical (Nash) equilibrium in the game, i.e. the best choice for all players is to stick to it. For \gls{pow} crypto ledgers, it is the "longest chain" rule. 

We aim to design a similar system, which also has an optimal strategy rule but which \textbf{does not require PoW competition}. We start from a broad perspective to derive the most generic principles of permissionless \gls{blockchain} design from Bitcoin and then apply these fundamental concepts to the Proxima architecture.

A \textit{ledger update} is defined as a \gls{deterministic} and atomic prescription that transforms a particular \gls{ledger state} $S_{prev}$ into another ledger state, $S_{next}$. The term \gls{deterministic} implies that the transformation always leads to $S_{next}$; the process is not affected by other ledger updates that may occur before, after, or in parallel. This transformation from $S_{prev}$ to $S_{next}$ can also be called a \textit{state transition}. An entity that produces state updates is said to be \textbf{writing to the ledger}. The term \textit{atomic} indicates that the ledger update is applied in its entirety to the ledger state, never partially.

In Bitcoin and other blockchains, ledger updates take the form of blocks. Each block consists of an ordered set of transactions and a block header. By design, each block is deterministic and atomic.

Ledger updates (blocks) are subject to \textbf{validity constraints} (or \textbf{ledger constraints}). Validity constraints refer to a publicly known algorithm that, when applied to the block, provides a \gls{deterministic} answer: either \textit{yes} for a valid block or \textit{no} for an invalid one. The assumption is that the absolute majority of participants in the system consider only valid blocks and reject invalid ones immediately. The participants enjoy full freedom of what ledger updates to produce as long as they do not violate ledger constraints. 

The validity constraints of a block typically involve verifying the validity of transactions within the block and checking the validity of the block header. In PoW blockchains, this includes verifying the PoW \textit{nonce}, which is easy to check but difficult to produce. The criteria for transaction validity in a block are usually blockchain-specific.

A block typically includes a \textbf{reward} for the issuer in the form of native tokens on the ledger, often newly minted (causing supply inflation), or collected as fees. The rewarding rules are subject to the ledger constraints, such as the halving rules in Bitcoin.

In the blockchain, different blocks at the same height serve as competing alternatives as ledger updates, requiring each node to decide which of several conflicting blocks to apply to its local copy of the ledger state. Ledger updates are freely broadcast to all participants, via the \textit{gossip protocol}, native to peer-to-peer systems. Each participant may choose any of them to apply to the ledger they maintain locally.

This setup leads to \textbf{competition} among block producers for their blocks to be accepted as updates to the ledger state. The \textbf{behavioral assumption} is that block producers will seek profit in the form of rewards, incentivizing them to win the competition. Given the high difficulty of PoW and the randomness of the nonce-mining process, the optimal strategy for profit-seeking block producers is to follow the longest chain. If they do not, their chances of catching up with the network diminish exponentially with each block they fall behind, decreasing their chances of receiving rewards. This is the well-known ``longest chain wins`` rule.

This results in a converging consensus-seeking behavior (also known as \gls{probabilistic consensus}) among the \textbf{permissionless, unbounded, and generally unknown set of writers to the ledger}\footnote{note that the \gls{probabilistic consensus}, strictly speaking, is not even a consensus in a classical \gls{bft} consensus sense. The latter assumes a bounded set of writers to the ledger (the \textit{committee}). Each consensus participant knows all other participants and is subject to a strict communication protocol, which is how opinions are shared and voted between them. Meanwhile, the Nakamoto consensus is more a "strategy" or "behavior" than a "protocol"}.

For us, the important general observation is the following:
\begin{observation}
 The "longest chain wins" consensus rule, followed by nodes, is not a primary fact or axiom, but rather derived from basic facts and assumptions as an optimal profit-seeking strategy that leads to a game-theoretical (Nash) equilibrium among participants. These base prerequisites include:
\begin{itemize}
    \item Specific validity constraints of the ledger updates, enforced system-wide.
    \item The \gls{liquidity} of the native token in the ledger, which incentivizes participants to seek profit.
\end{itemize}
\end{observation}

In our opinion, the distinguishing characteristic of the \textbf{Nakamoto consensus} is its open participation and permissionless writing to the ledger, rather than a specific consensus rule, which is derived from a particular set of ledger validity constraints.

We can view the validity constraints of Bitcoin blocks, PoW nonce in particular, as purposefully designed by Satoshi Nakamoto to achieve stable profit- and consensus-seeking behavior from otherwise independent and selfish writers to the ledger, in a hostile and unreliable environment created by the shared interest in the distributed ledger. The Bitcoin protocol enables \textbf{coordination of unwritten real-world social consensus among free pseudonymous participants} through purposely designed rules of the game. 

\begin{observation}
    Designing a permissionless \gls{crypto ledger} means designing the rules of a game: ledger validity constraints.
\end{observation}

%============================================================
\subsection{Proxima approach. From tangle to UTXO tangle}
\label{sec:proxima}

In Proxima, our benchmark is Bitcoin. Our goal is to design a \gls{crypto ledger} "game" for a fully permissionless set of writers to the ledger, avoiding the \gls{bft} setup with committees while employing a \gls{DAG}-based structure of \gls{ledger} rather than \gls{blockchain}.

We aim to design a set of ledger constraints in which UTXO \glspl{transaction} acts as a ledger update instead of blockchain blocks. The profit-seeking behavior of independent producers of UTXO transactions, enforced by the validity constraints, should converge to a consensus on the \gls{ledger state}. Similarly to the \textit{longest chain rule}, the \gls{biggest ledger coverage} rule emerges from the validity constraints as an optimal strategy leading to the Nash equilibrium among the participants in the consensus.

It is important to note the fundamental difference between blockchain blocks and UTXO transactions as ledger updates: 
\begin{observation}
Blockchain blocks, as ledger updates, always conflict with each other, placing block producers in constant \textbf{competition}. In contrast, UTXO transactions treated as deterministic ledger updates may or may not conflict, allowing multiple UTXO transactions to update the same ledger state in parallel, creating different yet non-conflicting parallel ledger states. These states can be consolidated and treated as one, facilitating \textbf{cooperation}.
\end{observation}

The main idea of cooperative behavior is drawn from the tangle. Let us quote the original \textit{The Tangle} paper:

\centerline{
\begin{minipage}{0.9\linewidth}
\textit{``The transactions issued by nodes constitute the site set of the tangle graph, which is the ledger for storing transactions. The edge set of the tangle is obtained in the following way:
when a new transaction arrives, it must \textbf{approve two previous transactions}. These approvals are represented by directed edges [...]``\\
``The transactions issued by nodes constitute the site set of the tangle graph, which is the ledger for storing transactions. The edge set of the tangle is obtained in the following way:
when a new transaction arrives, it must \textbf{approve two previous transactions}. These approvals are represented by directed edges [...]``
}    
\end{minipage}
} 
~\\
Furthermore: \newline

\centerline{
\begin{minipage}{0.9\linewidth}
    ``\textit{The main idea of the tangle is the following: to issue a transaction, users \textbf{must work to approve other transactions}. Therefore, \textbf{users who issue} a transaction are contributing to the network's security. It is assumed that the nodes check if the approved transactions are not conflicting. If a node finds that a transaction is in conflict with the tangle history, the \textbf{node will not approve} the conflicting transaction in either a direct or indirect manner}``
\end{minipage}
}
~\\
The principle of one transaction approving two previous transactions results in a DAG-like data structure that is constructed incrementally by adding new vertices. \gls{DAG} edges represent approval links. The specific number of approved transactions (two) is not essential; we generalize it to "any number from 1 to a fixed $N$".

\begin{wrapfigure}{H}{0.25\textwidth}
    \centering
    \includegraphics[scale=0.7]{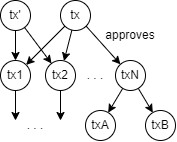}
    \caption{Tangle principle}
    \label{fig:tangle}
\end{wrapfigure}

We consider the principle of \textit{one transaction approving two previous transactions} as the foundation for \textbf{cooperative behavior}, as each ledger update consolidates several non-competing ledger updates into one history, unlike in the blockchain, which chooses one of many competing paths.

It is important to note that the original concept of the tangle was not specific about the ledger model it used, nor how the ledger relates to the tangle DAG. It is also ambiguous about which entities can add new transactions (vertices) to the tangle. In this context, it mentions users (token holders) and nodes, but these are different categories of entities, with only token holders possessing private keys required to sign a transaction.

We aim to fill these gaps by extending the generic tangle to the \gls{UTXO tangle}. The well-known UTXO ledgers (Bitcoin, Cardano) already form a DAG, with vertices as transactions and edges representing consumed outputs of other transactions. We expand this model by allowing optional \gls{endorsement} links to be added to the UTXO transaction. These \textit{endorsement} links resemble both the original tangle vertex approval links and the references of the blockchain block to the predecessor blocks. \textbf{Endorsement link indicate a declared and enforced consistency of the ledger update with the referenced ledger updates}.

These transactions form a DAG with two types of edges: (a) \textbf{consuming} the output of other transactions, and (b) \textbf{endorsing} other transactions. This data structure is known as the \gls{UTXO tangle}.

Endorsements are an atomic part of \gls{transaction}, signed by the private key of the same entity that moves tokens. Their position in the DAG is fixed by the \gls{token holder}; the transaction cannot be "re-attached" by anyone else.

\begin{figure}[ht]
    \centering
    \includegraphics[scale=0.7]{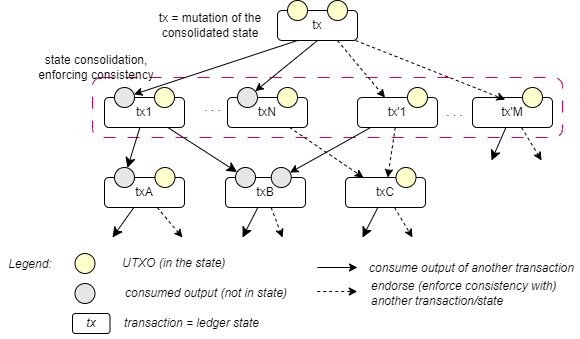}
    \caption{UTXO tangle}
    \label{fig:utxotangle}
\end{figure}

Here is our interpretation of the UTXO tangle (see figure \ref{fig:utxotangle}):  
\begin{enumerate}[(a)]
    \item Each vertex of the UTXO tangle represents a UTXO \gls{transaction} that \textbf{consumes} outputs from other vertices/transactions and may optionally \textbf{endorse} a chosen set of other vertices. 
    \item Each vertex represents a separate \gls{ledger state} resulting from the mutation of the consolidated state. The consolidated state is consistent among all inputs of the transaction, including endorsed ledger states.
    \item The UTXO tangle represents multiple ledger states. Any two ledger states (vertices) in the UTXO tangle may or may not be a \gls{conflict}. The \gls{past cone} of each vertex is always conflict-free. It represents the history of ledger updates that leads to the ledger state defined in a vertex.
    \item By adding a vertex to the UTXO tangle, another version of \gls{ledger} is created, a new ``fork`` of the state. Thus, the growth of the UTXO tangle involves the creation of forks and the consolidation of them.  
\end{enumerate}

In the UTXO tangle, we are specific about who can create a new vertex and add it to the tangle: the \gls{token holder} and no one else. The \textbf{token holders are the sole writers to the ledger}. Thus, each vertex of the UTXO tangle bears the identity of the issuer in the form of a signature on the vertex data.

%============================================================================
\subsection{Cooperative consensus principle}
\label{sec:cooperative}

The main principle of \gls{cooperative consensus} is the \gls{biggest ledger coverage} rule. The \glspl{token holder} are constantly maximizing the amount of capital involved in their transactions. This is analogous to the \textit{longest chain rule}, when \gls{pow} miner constantly maximizes the height of the block it produces. 

We introduce a special metric of the transaction (a vertex) on the UTXO tangle called \gls{ledger coverage} (or just \textbf{coverage}). 

\begin{definition}[\Gls{ledger coverage}]
\label{def:coverage}
The ledger coverage is calculated deterministically based on the past cone of the transaction, defined as the sum of the amounts of outputs that are consumed directly or indirectly by the past cone of the transaction within the specified \gls{baseline state}.
\end{definition}

Ledger constraints guarantee that the baseline state is always deterministically defined in the context of the transaction for which ledger coverage is calculated; i.e., in the valid UTXO tangle, each transaction has its baseline state. 
 
\begin{wrapfigure}{R}{0.5\textwidth}
    \centering
    \includegraphics[scale=0.6]{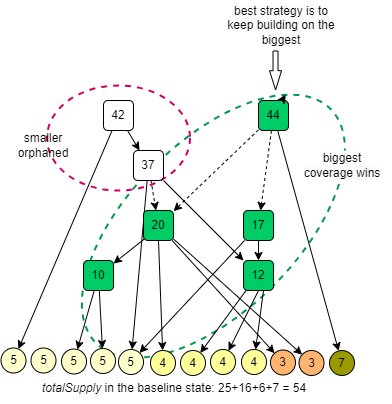}
    \caption{Ledger coverage explained}
    \label{fig:coverage}
\end{wrapfigure}

The \textit{ledger coverage} reflects the amount of funds "moved" by the past cone of the transaction in the baseline state. See the precise definitions in Sections \ref{sec:utxo-tangle}, \ref{sec:branches} and Figure \ref{fig:coverage}.  

Upon seeking the cooperative consensus:
\begin{enumerate}[(a)]
    \item Token holders, the sole entities participating in the protocol, exchange transactions. Each token holder builds their own UTXO tangle based on the received transactions. Transactions are not guaranteed to be received by a participant in the order of their creation and may be subject to arbitrary communication latency. 
    \item UTXO tangle can contain \glspl{conflict}: conflicting transactions and conflicting \glspl{ledger state}. Token holders share a common interest in reaching a consensus on the state of inclusion of transactions. In case of conflicts, different token holders ultimately prefer one conflicting transaction over others in the same way. This leads to consensus. 
    \item The transaction may include a reward as new tokens (the supply \gls{inflation}) or other tokenized form of interest. The \gls{liquidity} of the native token in the ledger drives the process toward profit-seeking behavior: token holders have economic interest in their transaction being included in the final ledger state, i.e., preventing their transaction from becoming \gls{orphaned}.  
    \item The token holders want their transaction to cover as much funds moved in the past cone as possible, according to the \gls{biggest ledger coverage} rule. The amount of capital involved in the ledger update indicates the vested interest the cooperating token holders have in that ledger update to become part of the final one. 
    \item By trying to maximize \gls{ledger coverage} of the new transaction, the token holder aims to make the transaction more attractive to other token holders to consolidate their transactions with the one produced.  
    \item The token holder has a random and asynchronously changing set of possible choices for consumption and endorsement targets, as new transactions from other token holders arrive unpredictably. The token holder must select a non-conflicting subset of all possible inputs to produce a transaction.
    \item The token holder aims to create the best possible outcome for the new transaction. In this way the choice of consolidation targets becomes a dynamic \textbf{optimization problem}, constantly aiming to find a non-conflicting set of inputs that maximizes an objective function, i.e. finding inputs with the largest possible \gls{ledger coverage}.
\end{enumerate}

Even without considering the continually changing set of possible choices, the optimization may be a resource consuming problem. The UTXO tangle graph is built in real-time, it has unbounded depth, and potentially contains many conflicting vertices. Each new transaction must contain conflict-free \gls{past cone}. To adhere to real-time constraints, even with relatively low transactions per second (TPS), practical application requires heuristics. Moreover, the token holder has only a limited time to construct the new transaction; falling behind the natural growth of the UTXO tangle increases the risk of the transaction being left outside the consolidated ledger and becoming \gls{orphaned}.

The \gls{ledger coverage} metric also represents a heuristic of how close the transaction is to the desired ledger state of the global consensus. We use this metric as the objective function of the optimization problem the token holder will normally be solving within a fixed time interval called a \gls{slot}. The maximum value of the ledger coverage has an upper bound, a constant (we skip details here). Intuitively, several non-conflicting ledger states with maximal ledger coverage (i.e., with all baseline outputs consumed) will be identical. Therefore, the growth of the ledger coverage suggests the convergence of multiple ledger states to a consensus ledger state. See below fig. \ref{fig:coverage} and section \ref{sec:convergence}.\nameref{sec:convergence}.

\begin{conclusion}
The \gls{token holder} aims to issue its new transaction with as large \gls{ledger coverage} as possible within time constraints. This is known as the \gls{biggest ledger coverage} rule.

The Nash equilibrium among token holders emerges from the fact that deviating from this strategy increases the chances that the transaction becomes less attractive to other token holders and therefore will less likely be included in the past cones of upcoming transactions which, in turn, will increase the chances of the transaction becoming \gls{orphaned}. 
\end{conclusion}

There is a guarantee that, within the boundaries of one slot and until maximal coverage is reached, a non-empty set of token holders can keep advancing with increasing ledger coverage by bringing their tokens to the UTXO tangle with new transactions. This process will continue as long as these token holders are incentivized to create new transactions.

The process is \textbf{cooperative} because only the token holder who controls the tokens can bring the corresponding ledger coverage to the UTXO tangle. Thus, each token holder can only ensure growing coverage for their transaction by consolidating the states of other token holders, i.e. by cooperating with them.

The \gls{cooperative consensus} is a \gls{Nakamoto consensus} by our definition, because it occurs among an unbounded and permissionless set of profit-seeking writers to the ledger. These entities make decisions without relying on global consensus on stakes/weight distribution of voting entities, a committee of voters/quorum, stakes, a coordinator, or similar mechanisms. They adhere to the \gls{biggest ledger coverage} rule as their optimal strategy, which is computed from their local copy of the UTXO tangle.  

%============================================================================
\subsection{The big picture}
\label{sec:big}

\begin{wrapfigure}{R}{0.42\linewidth}
    \centering
    \includegraphics[scale=0.42]{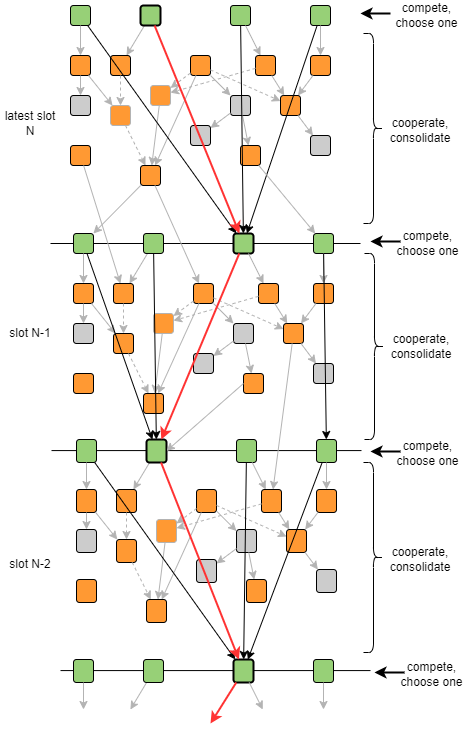}
    \caption{Cooperative consensus}
    \label{fig:big}
\end{wrapfigure}

The principle of \gls{cooperative consensus} along with the \gls{biggest ledger coverage} rule, must be enriched with many details and constraints to be applicable in an ever growing \gls{UTXO tangle}. Admittedly, the design becomes significantly more complex when compared with the genius simplicity of the Bitcoin design. This is the price we must pay when we include \gls{DAG} instead of \gls{blockchain} in the equation. In this section, we skip many details and present a broad overview of the protocol.

In our design, all \glspl{token holder} are allowed to build special structures, called \glspl{chain} (of transactions) in the ledger. During the construction of a chain, the token holder generates \gls{inflation} proportional to the amount of tokens held on \gls{chained output}. This is the fundamental incentive for capital holders to actively build chains instead of being passive.

A special permissionless class of active token holders, known as \glspl{sequencer}, plays a distinct role in the system. Each sequencer builds a chain of special transactions called \glspl{sequencer transaction}, each transaction consolidating as much \gls{ledger coverage} as possible. During \gls{slot}, all sequencers work together to cover \gls{baseline state} of the previous slot as best they can, given the time, performance, and communication delay constraints. A sequencer's success depends on its performance characteristics: its connectivity to other sequencers and the amount of on-ledger capital it can bring and offer as ledger coverage to other sequencers.

Non-sequencer token holders have their own strategies. They can \gls{tag-along} their transactions with the sequencer chains of their choice, offering a small or even zero fee (a "bribe") to incentivize the sequencer to consume the \gls{tag-along output}. The incentivized sequencer will "pull" the transactions to the future ledger state. 

The \gls{inflation} mechanism and the principle of \gls{tag-along} make sequencing a profit-seeking activity.

Large token holders can also make their tokens to participate in chain building by sequencers through \gls{delegation} thus receiving \gls{inflation} rewards in return. "Lazy whales" ("hodlers"), who do not delegate their holdings to some sequencer, face the opportunity cost of not earning inflation.

At the slot boundary, sequencers are given an opportunity to produce a special kind of sequencer transaction called a \gls{branch transaction} or \textbf{branch} to enter the next slot and win the random \gls{branch inflation bonus} (see section \ref{sec:bias}). Each branch transaction must select one branch transaction from the previous slot and consume its special output, known as the \gls{stem}. As a result, all branches on the same edge of the slot become \glspl{conflict}  and only one can be included in the ledger state. These constraints aim to guide the UTXO tangle towards a tree-like structure of branches and, ultimately, to a single path of branches, with the side branches quickly becoming \gls{orphaned}.

\textbf{At the slot boundary, sequencers compete rather than cooperate}. However, in order to be competitive on the boundary of the slot, all of them must cooperate during the slot.

The ledger states corresponding to the branches are committed to the databases maintained by the nodes connected to the network of peers. 

\subsection{Characteristics of the architecture}
\label{sec:characteristics}

To conclude the Introduction, we will highlight general characteristics of the proposed architecture.

\textbf{\Gls{permissionless decentralization}}. Fully permissionless consensus and writing to the ledger. \Glspl{token holder} can issue ledger updates without requiring any permitting procedures or voting.

\textbf{\Glspl{token holder} is the only category of participants in the consensus}. The architecture does not need special type of participants, like miners or validators, responsible for consensus. Raw transactions are the only type of message exchanged between consensus participants. The network of nodes plays the role of the technical infrastructure for the interaction of token holders.

The \textbf{\gls{cooperative consensus}} is based on the assumption of \gls{liquidity} of the token and on the profit-seeking behavior of \glspl{token holder}.

\textbf{Leaderless determinism}. The system operates without a consensus leader or block proposers. From the point of view of the \glspl{token holder}, \glspl{transaction} are \gls{deterministic}: the end state of the user assets is fully determined by the transaction.

\textbf{Partial asynchrony}. The architecture relies on weak assumptions about time: 
    \begin{itemize}
        \item The presence of a global time reference is assumed. Node operators and token holders are incentivized to maintain approximate synchronization of local clocks with the global reference; greater accuracy is preferable but not strictly required;
        \item prevention of certain attacks (solidification) requires assumptions about time bounds of transaction arrival.
    \end{itemize}

There are \textbf{no global registries} in the system, such as the list of participants, their stakes or rights. From the point of view of \gls{token holder}, there is no global account in the ledger, but instead a \gls{non-deterministic} set of controlled assets - \glspl{UTXO}.

Simple, \textbf{1-tier trust assumptions}. Only vested token holders are involved in the process, as opposed to the multiple layers of trust required in other blockchain systems such as \gls{pow} (which includes users and miners) and \gls{pos} (which includes at least users, block proposers, validators, committee selection mechanism, committee).

\textbf{Parallelism at the consensus level}: assets converge to their final state independently. 

Massive \textbf{parallelism at the node level}. All transactions are validated in parallel at each node.

%============================================================================
\section{UTXO tangle} 
\label{sec:utxo-tangle}

This section presents a more rigorous definition of \gls{UTXO tangle} and its constituent elements.

%============================================================================
\subsection{UTXO transaction} 
\label{sec:utxotx}

\begin{wrapfigure}{R}{0.5\textwidth}
    \centering
    \includegraphics[scale=0.6]{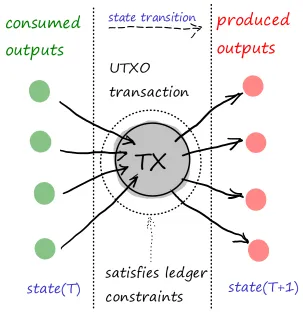}
    \caption{Classical UTXO transaction}
    \label{fig:utxotx}
\end{wrapfigure}

The Proxima model of \gls{UTXO ledger} is fundamentally similar to the well-known UTXO models of Bitcoin, Cardano (\mbox{EUTXO})\footnote{The Extended UTXO Model: \url{https://omelkonian.github.io/data/publications/eutxo.pdf}}, and others. However, it extends existing UTXO models with \glspl{endorsement} as a novel feature of the UTXO transaction, without altering the core principles.

The main concept of the UTXO ledger model is \gls{output} or \gls{UTXO}. The set of all outputs (UTXOs) constitutes \gls{ledger state}. Each output $O$ is uniquely identified in the ledger state by its \textbf{output ID} and has two mandatory properties: \textbf{amount} and \textbf{lock}. The \textit{amount}, denoted as $amount(O)$, represents the quantity of fungible tokens, native to the ledger, locked in the output $O$. The \textit{lock} specifies the conditions necessary to unlock the output, allowing it to be \textit{spent} or \textit{consumed}. The most common lock is target address (hash of public keys corresponding to secret private keys)\footnote{Other examples of locks in the Proxima ledger include chain locks, stem locks, delegation locks, tag-along locks, and more.}.

Each \gls{transaction} $T$ is composed of:
\begin{enumerate}
    \item An ordered set of outputs created by the transaction: $out(T) = (O^T_0, \dots, O^T_{n-1})$, referred to as \textbf{produced outputs} of the transaction $T$. The producing transaction \textit{ID} and output index constitute the \textit{output ID}, which is unique in \gls{ledger state};
    \item An ordered set of outputs spent by the transaction, termed \textbf{consumed outputs}: $in(T) = (O^{T{a}}_{i_a}, \dots, O^{T{z}}_{i_z})$. The IDs of these consumed outputs are also known as \textit{inputs} of the transaction $T$. The notation $T' \leftarrow T$ denotes that the transaction $T$ consumes some output of $T'$. In this case, we also say that $T'$ is \textit{parent} of $T$, and $T$ is \textit{child} or \textit{descendant} of $T'$;
    \item Optionally, up to $N$ endorsements of other transactions: $endorse(T) = (T_0, \dots, T_k), k < N$. The notation $T' \Leftarrow T$ denotes that transaction $T$ endorses transaction $T'$. Here, $N$ is the global constant;
    \item A \gls{timestamp}, representing the \gls{ledger time} of the transaction denoted as $timestamp(T)$. This will be discussed in section \ref{sec:nodes}. \nameref{sec:nodes}.
    \item contains \textbf{input unlock data} denoted as $unlock(T)$;
    \item contains cryptographic \textbf{signature} (including public key) of all the data above. Ensures that all the above data are approved by the \gls{token holder}. It is denoted as $senderID(T)$; 
    \item Optionally, other \textbf{auxiliary data}.
\end{enumerate}

Therefore, a transaction $T$ is a tuple: 
$$
(in(T), out(T), endorse(T), timestamp(T), unlock(T), senderID(T))
$$ 
The \textit{transaction ID}, denoted as $id(T)$, is a deterministic function of all the transaction data listed above. Usually, it includes the value of a one-way hash function. $id(T)$ uniquely identifies the transaction $T$ on the UTXO tangle.

A transaction is considered valid if it is syntactically correct and satisfies all validity constraints imposed on the consumed and produced outputs, timestamp, endorsements and on the transaction's \gls{past cone}.

%============================================================================
\subsection{Transaction graph. Past cone}
\label{sec:past}
Consider a set of transactions $L=\{T_0, \dots, T_n\}$. 
Let $L$ be a vertex set of a graph. Let's also assume the edge set of the graph consists of all pairs of transactions from $L$, that are connected either by a consumption relation $\leftarrow$ or endorsement relation $\Leftarrow$: $$
edges(L)= \{(T_a, T_b) \| T_a \leftarrow T_b \lor T_a \Leftarrow T_b\}
$$ 
This naturally forms a \textit{directed graph} in the set of transactions $L$. 

For simplicity sake and when context allows, we will use $L$ (or other vertex sets) for both the vertex set and for the graph $(L, edges(L))$ itself interchangeably\footnote{in this paper we will relax requirements for strict mathematical notations whenever meaning can be derived from the context. In particular, each transaction $T$ is a data structure, and the edges of the graph are one-way hash functions of transactions. So, we can often skip referencing particular sets of transactions because each constituent element of the transaction graph corresponds to an element in the global set of all possible values of the hash function and collisions are practically impossible}.

We assume that the directed graph $L$ is also \textit{connected} and \textit{acyclic}, i.e. it is a \gls{transaction DAG}. Then \gls{past cone} of the transaction $T \in L$ we define as a subset $past(T) \subset L$, where each transaction $T' \in past(T)$ is reachable through the relations $\leftarrow$ and $\Leftarrow$ between the vertices.

We say that two distinct transactions $T_a, T_b \in L$ constitute \gls{double-spend} or \gls{conflict}, if there's an output, that is consumed by both $T_a$ and $T_b$, i.e., if their consumed outputs intersect: $$
in(T_a) \cap in(T_b) \neq \emptyset
$$ 
A set of transactions $L$ is \textit{consistent}, if it does not contain a double-spending pair of transactions. Similarly: 
\begin{enumerate}
    \item Any two transactions $T_a, T_b \in L$ constitute \gls{conflict}, if union of their past cones contains a \gls{double-spend}.
    \item A set of transactions $\{T_a, T_b, \dots \} \subset L$ is \textbf{consistent}, if union of their \glspl{past cone} does not contain a conflicting pair of transactions.
    \item A transaction $T \in L$ has valid \gls{past cone}, if $past(T)$ is consistent.
\end{enumerate}

%============================================================================
\subsection{Ledger. Baseline state}
\label{sec:ledger}
Any consistent set of transactions $L = \{T_1, \dots, T_n\}$ is called a \gls{ledger}. Each ledger $L$ has a corresponding set of \glspl{output} (UTXOs) denoted $S_L = \{O_1, \dots, O_m\}$, known as \gls{ledger state}.

Usually, each output in the ledger state is associated with the transaction that produced it. As an exception, we identify a set of outputs in $L$, called \gls{baseline state}, which does not have corresponding producing transactions in the ledger. This set of outputs is denoted as $G$ or $G_L$. A valid ledger can be constructed incrementally from a baseline state $G$ by adding transactions.

Any consistent ledger state may serve as a baseline for developing a ledger on top of it. We can define a valid ledger recursively:
\begin{itemize}
    \item The empty set of transactions is a valid ledger with a corresponding ledger state equal to the baseline state $G$: $S_{\emptyset} = G$.
    \item Suppose $L$ is a valid ledger with the ledger state $S_L$. We can append a new transaction $T$ to the ledger if:
    \begin{itemize}
        \item All consumed inputs exist in the ledger state, i.e. $in(T) \subset S_L$
        \item All endorsed transaction are on the ledger, i.e. $endorse(T) \subset L$
        \item The transaction $T$ is \textit{valid}
        \item The resulting set of transactions $L \cup \{T\}$ is consistent (meaning that the consolidated past cone is free of conflicts/double spends)
    \end{itemize}
    In this case, transaction $T$ is said to be \textbf{applicable} to the ledger state $S_L$.
    \item The new ledger $L'= L \cup \{T\}$ is a valid ledger by definition. The corresponding ledger state $S_{L'}$ of the updated ledger $L'$ is constructed by removing consumed outputs from $S_L$ and adding all produced outputs to it: $S_{L'} = S_L \setminus in(T)\cup out(T)$
\end{itemize}
The valid ledger $L$ and its state $S_L$ is the one which can incrementally be constructed from the baseline ledger state $G$ by adding valid transactions, that are consistent with the existing ones. This construction ensures that the ledger $L$ is always connected and acyclic, i.e. a \gls{transaction DAG}. Note that each valid ledger is free of double consumption of outputs. i.e. it is consistent.

By definition, the genesis outputs contain all tokens initially defined in the ledger, and the sum of the amounts equals the total initial supply of tokens\footnote{strictly speaking, total supply is a function of time due to the inflation rules. This fact is not essential in the context, so we assume it is constant}:
$$
totalSupply = \sum_{O \in G} amount(O)
$$

We can also look at the transaction from the \textbf{ledger state update} perspective. Let us say transaction $T$ is \textit{applicable} to the ledger state $S$. We can see $T$ as the following atomic transformation of the ledger state $S$ to the ledger state $S'$:
\begin{itemize}
    \item deleting consumed outputs $in(T)$ from the ledger state
    \item adding produced outputs $out(T)$ to the ledger state
\end{itemize}

In this way, transaction $T$ will transform state $S$ to $S'$. We can denote the resultant ledger state $S' = S^T$.

\begin{observation}
Two transactions $T$ and $T'$ with consumed sets that do not intersect ($in(T)\cap in(T')=\emptyset$) in the ledger state $S$ will be \textbf{independent and parallel}. The resulting ledger states $S^T$ and $S^{T'}$ will be non-conflicting and can naturally be consolidated into one ledger state $S^{T,T'}$. This trait of the UTXO ledger is crucial for \textbf{parallelism of transaction validation and consensus}.
\end{observation}

\begin{observation}
The incremental construction of \gls{UTXO ledger} form its \gls{genesis ledger state} is \gls{deterministic}: final result does not depend on the exact order in which transactions are added to it, akin to a jigsaw puzzle.
\end{observation}

%======================================================================
\subsection{Definition of UTXO tangle. Multi-ledger}
\label{sec:multi-ledger}
\noindent 

\begin{definition}[\gls{UTXO tangle}]
Let $U = \{T_1, \dots, T_N\}$ be a finite set of transactions represented as vertices, where \gls{past cone} of each transaction $T \in U$ traces back to a common genesis $G$. Although the set of transactions $U$ itself may or may not be consistent, by definition we assume that the past cone of each individual transaction $T \in U$ constitutes a consistent ledger $L_T$. The UTXO tangle is formally defined as the transaction DAG naturally induced by the set $U$.    
\end{definition}

So, each vertex of the UTXO tangle represents a consistent ledger; however, \gls{UTXO tangle} may contain \glspl{conflict} in the consolidated past cone. In such cases, the corresponding consolidated past cones would be conflicting, and resulting ledger states will not be represented as transactions. 

Thus, a \gls{UTXO tangle} is a data structure that may contain multiple versions of ledgers, both conflicting and non-conflicting. This trait is known as \textbf{multi-ledger}.

%======================================================================
\subsection{Transaction validity constraints}
\label{sec:constraints}

Each transaction is subject to enforced validity rules that we denote as a value of the predicate $valid(T)$. By definition, the validity of the transaction is equivalent to the consistency of the ledger $L_T$ and the corresponding ledger state $S_{L_T}$. 

So, in our definition, we include the validity of \gls{past cone} in the validity of the vertex, the transaction. Therefore, the predicate $valid(T)$ is a function of $T$ itself and its past cone, the history of the ledger state. 

Transactions always have a consistent past; otherwise, they are not transactions and cannot be added to the UTXO tangle. 

%======================================================================
\subsubsection{Validity of the past cone}
Validity of a transaction (a vertex) in the UTXO tangle includes requirement that its past cone is both (a) conflict-free (refer to \ref{sec:past}.\nameref{sec:past}) and (b) forms a \textit{connected graph} (also known as being \textit{solid}). This in particular means that each transaction is connected with the genesis by the path of consumption or endorsements relations.

%======================================================================
\subsubsection{Transaction level validity constraints}
By \textit{transaction level constraints}, we mean the constraints that do not depend on the entire past cone but cannot be attributed solely to individual outputs. These constraints include at least:

\begin{itemize}
    \item Checking syntactical transaction composition.
    \item Enforcing token balance between inputs and outputs, including \gls{inflation} constraints. 
    \item Verification of the validity of \gls{timestamp}: The timestamps of the inputs must be strictly before the transaction timestamp by a global constant called the \textbf{transaction pace}. 
    \item Validity of the signature.
    \item Validity of all consumed and produced outputs. See \ref{sec:validity-of-outputs}
\end{itemize}

%======================================================================
\subsubsection{Validity of outputs}
\label{sec:validity-of-outputs}
The absolute majority of transaction validity checks are expressed through the use of validation scripts, immutably attached to \glspl{output} in \gls{ledger state}. The UTXO is also said to carry a \textbf{covenant}, a kind of (smart) contract which is enforced by protocol between the owner of the output and the consumer of it.  

By definition, each output $O$ is composed of scripts called \glspl{constraint}\footnote{we use term \textit{constraint} to emphasize the fact that UTXO transactions, being \gls{deterministic}, are never "executed", but are "validated" instead. It is different from non-deterministic model of transactions, such as Ethereum, which are "executed".}: each output is vector of scripts $O=(c_1, \dots c_n)$. Each script $c_i$, when evaluated, results in \textit{true}, if the constraint is satisfied in the given evaluation context, or \textit{false} if it is not satisfied (violated)\footnote{The \textit{validation script} In Proxima is computationally equivalent to Bitcoin Script or similar engine. It is intentionally non-Turing complete. We will not delve into this deeply and broad topic here}.

The output $O$ is considered \textbf{valid in the validation context}, if all of its validation scripts $\{c_i\}$ render \textit{ true} when run in that context. 

Each output and each of its validation scripts can be evaluated in two different contexts: \textit{as produced output} and \textit{as consumed output}.
\begin{itemize}
    \item When evaluated as a \textit{produced output}, the script $c_i$ of output $O$ is run in the context of the producing transaction $T_{prod}: O \in out(T_{prod})$. In other words, the script is provided with the data of the producing transaction $T_{prod}$ as parameter, along with the output itself.
    
    \item When evaluated as a \textit{consumed output}, the script $c_i$ of output $O$ is run in the context of the consuming transaction $T_{cons}: O \in in(T_{cons})$. This means that along with the output itself, the script is provided with the data of the consuming transaction $T_{cons}$ as parameter.

\end{itemize}
A consumed output can be validated without requiring the corresponding transaction that produced it. This means that the inputs of the transaction can be validated independently of the past cone. 

Validation scripts of the consumed outputs typically involve checking their output unlock data, which must be provided by the consuming transaction. For example, it may include verification if the address data in the lock constraint match the signature $senderID(T)$ in the consuming transaction $T$.

The ledger definition contains a static library of standard scripts that are included in the ledger genesis. For example, the standard \textit{chain constraint} is a library script, \textit{addressED25519} lock script is a standard script, and there are many others.\footnote{Note, that each script is taking its parameter data from the context only. For example, running the script $addressED25519(0x97459a12301950535caa5115af5edbc04ead272c25880c967ee522dc11bc8193)$ in the consumed output means checking if the output which is locked in this particular target address is unlocked in the context of the consuming transaction}.

The scripting capabilities and the ability to compose different scripts as "Lego blocks" in the outputs make Proxima UTXO ledger a programmable UTXO ledger\footnote{In Proxima we use simple yet expressive functional language \textit{EasyFL}: \url{https://gitub.com/lunfardo314/easyfl} for output constraint scripting. Almost all transaction validity constraints, needed for the Proxima consensus on the UTXO tangle, are expressed as \textit{EasyFL} formulas}.

%=======================================================================
\section{Nodes} 
\label{sec:nodes}
We envision a system in which the only participating entities are \glspl{token holder}. Token holders produce transactions, exchange them with each other, build their local \glspl{UTXO tangle} and maintain consensus on the ledger state. In the Proxima narrative, the \textbf{distributed ledger is maintained by token holders rather than nodes}.

Of course, token holders (humans or organizations) require infrastructure support, and here comes \textbf{nodes} connected to the network. The nodes are not independent entities; they do not produce transactions, express opinions about transactions, or maintain consensus on the ledger state.

The primary function of the network of nodes is to provide a distributed infrastructure for token holders, who are responsible for achieving consensus on the ledger. The Proxima nodes do not serve the roles of a \textit{miners} (as in \gls{pow} systems) or a \textit{validator} (as in \gls{pos} systems).

The main functions of a Proxima node are as follows:
\begin{itemize}
    \item \textbf{Maintain Connections with Peers}: nodes establish and maintain connections with other nodes in the network.
    \item \textbf{Gossip}: nodes broadcast (relay) newly received transactions to their peers on the network.
    \item \textbf{Synchronize and maintain local UTXO tangle}: nodes synchronize with the network to maintain a valid copy of the UTXO tangle locally. They solidify and verify the validity of each transaction before adding it to their local copy.
    \item \textbf{Commit ledger states} of each branch transaction (see Section \ref{sec:branches}) to the local database. This function may include database cleanup and state pruning.
    \item \textbf{Provide read access} to \gls{UTXO tangle}, to \glspl{ledger state} of each branch for token holders (including \glspl{sequencer}) and other users. 
     \item \textbf{Provide stored raw transactions} to other nodes on request.
     \item \textbf{Synchronize local clocks}: The nodes synchronize their local clocks with the global time reference and the \textbf{ledger time} to ensure approximate synchronicity of local clocks across the network.
    \item \textbf{Protect against spamming attacks}: nodes are responsible for protecting themselves and their peers from potential spamming attacks to maintain network integrity (see \nameref{sec:anti-spam}).
\end{itemize}
The nodes communicate by exchanging raw transaction data: the main type of messages shared between nodes.

Token holders access the network by examining their local copy of the UTXO tangle, either from their own nodes or through public services. Token holders review their local UTXO tangle to make decisions about producing and sending transactions to the network. The nodes themselves do not have any intention or ability to issue transactions.

%=======================================================================
\section{Ledger time} 
\label{sec:ledger-time}

We assume the existence of a global world clock that serves as the time reference shared by all nodes in the network. This assumption is practical as long as the network remains on the planet Earth. We cannot guarantee perfect synchronization of local clocks across all nodes, but nodes are expected to keep their clocks reasonably close to the global reference time.

Node operators (token holders or those who profit from running public nodes) are incentivized to keep their nodes' clocks synchronized with the global clock as closely as possible. Significant deviations between the local clock and the global clock will lead to a higher risk of transactions being \gls{orphaned}, therefore, token holders are incentivized to keep (or choose) nodes synchronized with the global clock. We will elaborate on this in the discussion on \nameref{sec:sequencing}. 

The deviation of local clocks from the global one is expected to be symmetrically distributed around zero with small variance. The network can tolerate significant differences between local clocks of part of nodes, as long as the clocks of the largest token holders' nodes are roughly in sync.

The ledger uses a \textbf{logical clock} called \gls{ledger time}, expressed in \glspl{tick}. Each transaction $T$ includes a $timestamp(T)$, which is the number of \glspl{tick} since genesis. \Gls{slot} is equal to $128$ ticks.

The relationship between the ledger time scale and the real time scale is established by two static ledger parameters:
\begin{enumerate}
    \item \textbf{genesis time}, an absolute moment of the global clock\footnote{usually expressed as the so-called Unix time}
    \item \textbf{tick duration} on real time scale\footnote{In the testnet version, tick duration is set to 80 milliseconds, making the slot duration ~10.24 seconds}
\end{enumerate}

The correspondence between ledger time and real time is given by the formula:
$$
realTime(T) = genesisTime + timestamp(T) \cdot tickDuration 
$$
Slot of the transaction:
$$
slot(T) = \lfloor timestamp(T) / 128 \rfloor
$$
Ticks in the transaction:
$$
ticks(T) =  timestamp(T) \mod 128
$$

\begin{wrapfigure}{H}{0.6\textwidth}
    \centering
    \includegraphics[scale=0.6]{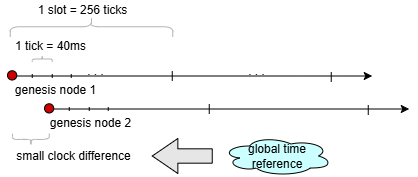}
    \label{fig:ledgertime}
    \caption{Local clocks and ledger time}
\end{wrapfigure}

The timestamp of the transaction is subject to validity constraints at the transaction level. The \textbf{timestamps of inputs and endorsed transactions must be strictly before the transaction timestamp}\footnote{This implies that sorting transactions by timestamp results in a topological sorting of the DAG} by at least a specified number of ticks, known as \textbf{transaction pace} constant. 

The node's responsibility is to synchronize the ledger time of incoming transactions with its local clock. This is achieved by \textbf{delaying} the processing of the transaction with a timestamp that translates to a value that precedes the node's current local clock time. The transaction is delayed until the local clock reaches the timestamp, ensuring that the node remains synchronized with the ledger time.

If a node with a local clock running ahead of schedule sends a transaction to a peer node with a local clock running behind, the receiving node will perceive the transaction as arriving from the future. Consequently, the transaction will be held in a "cool-down room" until it reaches its scheduled time according to the receiving node's local clock. This behavior is expected of the majority of nodes, ensuring an approximate synchronicity of local clocks with the ledger time throughout the network.

This behavioral assumption is considered realistic because producing a transaction either ahead of or behind the network's average can increase the chances that the transaction will be orphaned. It incentivizes token holders to keep their transactions in line with the network's time standards, ensuring better compatibility and synchronization across the network.

There is no strict need for the lower bound for \gls{ledger time} of incoming transactions, except in the situation of preventing certain types of spamming attacks\footnote{It may be practical to introduce a certain past time horizon for transactions too far back on the ledger time axis to be rejected by the node. In case of different reactions of nodes on the transaction due to difference of local clocks, the transaction can be requested if needed during solidification of the past cone}.

%=======================================================================
\section{Chains}
\label{sec:chains}

\begin{wrapfigure}{H}{0.4\textwidth}
    \centering
    \includegraphics[scale=0.6]{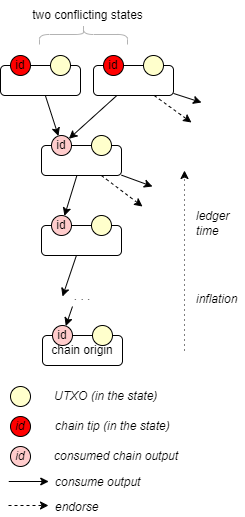}
    \caption{Chain}
    \label{fig:chain}
\end{wrapfigure}

\Glspl{chain} play a crucial role in the Proxima ledger and consensus. It is also a base for the incentives mechanism. 

\begin{definition}[\Gls{chained output}]
The chained output is any \gls{output} (UTXO) that has attached a standard validation script called \gls{chain constraint} to it. Each \textit{chain constraint} is tagged with a unique \textbf{chain ID}. The script with specific \textit{chain ID} in the output $O$ enforces the following validity constraint: output can be consumed in a transaction $T$ only if one of the following conditions is met:
\begin{enumerate}[(a)]
    \item The transaction produces exactly one \gls{chained output}, known as a \textbf{successor}, with the same \textit{chain ID}. In this case, the consumed output is referred to as a \textbf{predecessor}.  
    \item The chain is explicitly terminated, in which case the transaction does not produce any \textbf{successor} output.
\end{enumerate}

\end{definition}
\begin{definition}[\Gls{chain}]
The chain with a particular chain ID is a sequence of transactions on the UTXO tangle, where each next transaction in it produces the successor of some chained (produced) output in the previous one. 
\end{definition}

A transaction may belong to several chains. The chain has unlimited length. 

\Glspl{output} (UTXOs) in the UTXO ledger state are single-time transient assets that exist until they are consumed by a transaction. The \textit{chain} transforms a sequence of outputs into a permanent, non-fungible asset with potentially unlimited lifetime. This provides an important mechanism for continuity and permanence in the ledger. Thus, the chain constraint enforces a chain of outputs on the UTXO tangle, identified by the chain ID as a unique and persistent identifier for the chain's lifetime.

The corresponding chain-constrained output, the tip of the chain, is immediately retrievable in the \gls{ledger state} by the \textit{chainID}. This makes each chain a \textit{non-fungible asset}, which can be utilized for various purposes, such as implementing NFT-s. In Proxima, chains are primarily used to implement \glspl{sequencer milestone} and \glspl{branch transaction} (as discussed later).

Note that if two different transactions attempt to \textit{extend} the same \gls{chained output} to their respective successors, this will result in \gls{double-spend}. In other words, these transactions will represent two conflicting ledger states, creating a fork in the ledger. \textbf{Within one consistent ledger, chain forks are not possible}.

\section{Inflation. Incentives} 
\label{sec:inflation}
\noindent
\Gls{inflation} is increasing the total supply of tokens in the ledger. In Proxima, inflation serves as the basis for an incentive mechanism and is closely related to the concept of \glspl{chain}. This design aims to reward the participants in the network for maintaining the integrity of the ledger and promoting healthy economic dynamics within the ecosystem with the controlled generation of new tokens over time\footnote{current version of the inflation mechanism simplifies the one presented in the earlier versions of the whitepaper by removing thresholds and making it more linear and directly computable on-ledger with basic formulas of \textit{EasyFL}}.

As a general principle, we want inflation $I$ to be proportional to time and funds $A$:
$$
I = A\times R \times \Delta t
$$
where $\Delta t$ is the duration of time and $R$ is an inflation rate.

\subsection{Inflation definitions}
\label{sec:inflation-formulas}

To apply this general principle to the discrete time scale and specifics of the Proxima ledger, let us consider a chain transition (produced by a transaction) where the predecessor is a (chained) output $O$ and its successor is output $O'$.

Let us denote $A=amount(O)$ and $A'=amount(O')$. $I$ the number of new tokens created in this transition:
$$
A'=A+I
$$

Let $t$ be the slot of the output $O$ and $\Delta t$ be the number of full slots between $O$ and $O'$. We define $I_t$ inflation on the slot $t$ of the amount $A$ as follows:
\begin{definition}[Single-slot inflation]
$$
I_t = 
\begin{cases}
    0 & \Delta t = 0 \\
    R_t \cdot A & \Delta t \ge 0
\end{cases}
$$
Where
$$
R_t = \frac{1}{C+t}
$$
\end{definition}

$R_t$ is the inflation rate in the slot $t$. It declines with every slot. The constant $C$ defines how steep the decline is. 

The transition of a chain by a transaction generates zero new tokens if the predecessor and successor are in the same slot. The inflation amount is proportional to $R_t$ if it is in different slots.

Let us say $A_t$ is some amount in the chain output in the slot $t$.
The successor of the output will always produce inflation $I_t$ (one slot), no matter how long the gap between the outputs is. 

The sequence of $k$ chained outputs in $k$ subsequent slots will produce amount on the output:
$$
A_{t+k} = A_t\cdot(1+R_t)(1+R_{t+1})\dots (1+R_{t+k-1}) = A_t\cdot\prod_{i=t}^{t+k-1}(1+R_i)
$$

$$
1+R_i = \frac{C+i+1}{C+i} \text{ for }i>0
$$

$$
A_{t+k} = A_t\cdot\frac{\bcancel{C+t+1}}{c+t}\cdot \frac{C+t+1+1}{\bcancel{C+t+1}} \dots \frac{C+t+k-1+1}{\bcancel{C+t+k-1}} = \frac{C+t+k}{C+t}
$$

$$
A_{t+k} = A_t\cdot (1+k\cdot R_t)
$$
The formulas
$$
I_{A,t,k}=k \cdot \frac{A}{C+t}
$$ 
$$
A_{t+k} = A_t +  k \cdot \frac{A_t}{C+t}
$$
respectively gives inflation of the amount $A$ starting with the slot $t$ over the period of $k$ slots and the resulting number of tokens. 

\begin{wrapfigure}{H}{0.4\textwidth}
    \centering
    \includegraphics[scale=0.9]{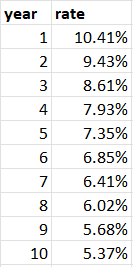}
    \caption{Inflation rate projections C=30,303,030}
    \label{fig:inflation-table}
\end{wrapfigure}

This is our intention: \textbf{inflation and the total amount are linear over time}.

We adjusted the formulas above for integer arithmetic. We can see that for inflation $\frac{A_t}{C+t}$ to be positive in integer arithmetic, we need the following condition satisfied:
$$
A_t \ge C+t
$$ 
The above is the \textbf{condition of minimum on-chain balance with non-zero inflation} where $C$ plays the role of \textbf{ minimum inflatable amount} of tokens in slot 0.

In figure \ref{fig:inflation-table}, we provide projected annual inflation values with $C=30,303,030$:

The balance validity check of input-output amounts for a transaction $T$ ensures that the sum of the output amounts produced does not exceed the sum of the input amounts plus the total inflation of all produced \glspl{chained output}. This check is essential to maintain the integrity of the ledger and to enforce the inflation limits set by the constraints. Given a transaction $T$, the validity check can be expressed as follows:
$$
\sum_{o \in out(T)}amount(o) \le \sum_{o \in in(T)} amount(o) + \sum_{o \in out(T)}inflation(o)
$$  

\subsection{Intended behavior}
\label{sec:intended-behavior}

We want to incentivize constant participation of token holders with their tokens in the consensus.

The enforced inflation value gives linear growth of the inflated tokens on the chain as long as the token holder issues at least one chain output per slot. 

Inflating the amount of their holdings is completely permissionless, and every token holder is entitled for the inflation rewards, while stopping issuing transaction for the token holder means implicit cost of dilution. Likewise, \gls{branch inflation bonus} encourages token holders to issue branch transactions.

Total inflation in the Proxima ledger is directly related to the activity of token holders, incentivizing participation and engagement in the network. This incentives mechanism functions through three main channels:

\begin{enumerate}[(a)]
    \item token holders can choose to run a \gls{sequencer}. By doing so, they proactively build the sequencer \glspl{chain} with inflation and potentially benefit from transaction fees.
    \item token holders earn inflation rewards through the \gls{delegation} mechanism. The \gls{delegation} effectively is lending tokens to support the \gls{sequencer}'s operations, while expecting a share of the generated inflation. The delegation mechanism is crucial for the scalability of cooperative consensus.
    \item token holders can \gls{tag-along} their transactions in a sequencer chain, aligning their activities with the sequencer's operations. This allows them to benefit from the sequencer's efficiency and potential inflation.
\end{enumerate}

Inflation is proportional to the total amount of capital that is moving around in the ledger. This model mirrors traditional financial concepts such as \textit{return on capital} and \textit{time-is-money}, where more active and engaged token holders are rewarded. Conversely, passive holders or "lazy whales" who do not actively participate in the network's operations do not receive inflation rewards. This creates an equitable system that encourages active participation in the security of the network.

We want to stress that the philosophy of Proxima does not encourage passive token holding behavior, common in most blockchains, where moving tokens usually means cost of fees. We want every token to participate in the security of the ledger. 

The fear of inflation, also common in crypto communities, does not have any grounds in Proxima. The chain \textbf{inflation is capped, predictable, and proportional for each token holder} (equity principle). The additional branch inflation bonus is available to sequencers so that they can offset their costs of running the sequencer nodes and bearing risks. 

We believe that deterministic universal inflation reflects common principles of how value markets function and this can be the basis for sound tokenomic of the Proxima ledger. 

%=======================================================================
\section{Sequencing} 
\label{sec:sequencing}

The Proxima ledger is entirely dependent on the activity and behavior of its token holders, making it a very democratic system.

It is up to \glspl{token holder} to choose what to do or not in the ledger, as long as their activity complies with the ledger constraints. We design Proxima as a pro-active profit-seeking game between token holders, where imposed rules (constraints) of the game lead them to cooperative and consensus-seeking behavior.

With designed ledger constraints, we want to encourage certain types of game strategies among token holders. Being a \gls{sequencer} is the main strategy to follow for large token holders, while contributing to the security of the ledger.   

\textbf{Sequencing} means building \glspl{chain} of \glspl{sequencer milestone} according to the \gls{biggest ledger coverage} rule. It plays a role similar to the block production performed by miners and validators in other distributed ledger systems\footnote{Sequencing has strong similarity with the sequencing for L2 chains/rollups, hence the name. The Proxima sequencer, as an option, can be a \textit{centralized or distributed sequencer} in the sense of L2 chains, so the name is not misleading}.

\subsection{Sequencers}
\label{sec:sequencers}

There is substantial evidence that distributed ledgers tend to converge towards a high concentration of capital no matter what, where a small number of entities control the vast majority of assets on the ledger\footnote{see for example \textit{How centralized is decentralized? Comparison of wealth distribution in coins and tokens}: \url{https://arxiv.org/abs/2207.01340}}. Prominent examples include major exchanges and token custodians which are entrusted to manage large amounts of on-chain assets. This appears to be a fundamental and unavoidable trend.

With Proxima, instead of worrying about "fairness" of the capital distribution or attempting to create a more equitable world, the design embraces practicalities. It assumes that a significant concentration of tokens in the hands of a few is normal and follows the power law, exhibited by many biological and social systems. Consequently, major token holders (often "wales" but primarily large exchanges and other custodians) will eventually command a substantial share of the capital, which entails a substantial amount of \textit{skin-in-the-game}, \textit{commitment}, and \textit{trust} in the distributed ledger (more about this in \ref{sec:social_consensus}.\nameref{sec:social_consensus}).

Proxima intentionally distinguishes a particular class of token holders known as \glspl{sequencer}, who are capable and willing to take on a significant portion of the shared ledger's responsibility in exchange for certain privileges compared to non-sequencers. A sequencer can be any entity that controls more than a minimum amount of capital. This minimum amount required for sequencing is defined as a constant in the genesis of the ledger\footnote{For example, in the testnet implementation, the minimum amount on the sequencer chain is set to 1/1000 of the initial total supply, allowing for a maximum of 1000 sequencer chains. The optimal value for this constant is an open question, particularly in the context of other ways of contributing to the capital of the sequencer, such as \textit{delegation}}

\begin{definition}[\Gls{sequencer milestone}]
A sequencer milestone is a \gls{chained output} with special \textbf{sequencer constraint} validation script attached to it.      
\end{definition}

\Gls{sequencer transaction} is a transaction that produces a sequencer milestone. A \textit{sequencer chain} is a chain composed of sequencer transactions. 

Any \gls{token holder} with sufficient capital can become a \gls{sequencer} by issuing sequencer transactions and maintaining persistent sequencer chains. There are no registration, no on-boarding, no staking, or other preliminary steps as seen in many \gls{pos} networks. This allows for a direct and straightforward approach to becoming a sequencer by its own decision without the need for any permissions.

Sequencers, by running sequencer chains, will generate returns on their own and leveraged capital in the form of \gls{inflation}, similarly to any other \gls{token holder} running a \gls{chain}. In addition to these returns, sequencers also have the opportunity to receive \gls{branch inflation bonus}, collect \glspl{tag-along fee}, and earn \gls{delegation} margins in exchange for their services to the network. These incentives provide sequencers with additional income streams beyond standard inflation, rewarding their active participation and support of the network.

Each \gls{sequencer milestone}, enforces additional constraints on a \gls{chained output}:
\begin{enumerate}[(a)]
    \item Only \glspl{sequencer transaction} are allowed to endorse other transactions, thus consolidating non-conflicting ledgers.
    \item Cross-slot \glspl{endorsement} is prohibited, that is, the endorsed transaction must be in the same slot.
    \item Only sequencers can create \glspl{branch transaction}. A sequencer transaction becomes a branch transaction when its \gls{timestamp} is on a \gls{slot} boundary (i.e., with $ticks(T) = 0$). Branch transactions represent \glspl{baseline state} for the other sequencer transaction in the slot. See Section \ref{sec:branches}.
    \item Each \gls{sequencer transaction} must have an \textbf{explicitly or implicitly referenced \gls{baseline state}}. See Section \ref{sec:stem} for more details.
\end{enumerate}

%==========================================================================
\subsection{Branches}
\label{sec:branches}

\begin{wrapfigure}{H}{0.4\textwidth}
    \centering
    \includegraphics[scale=0.45]{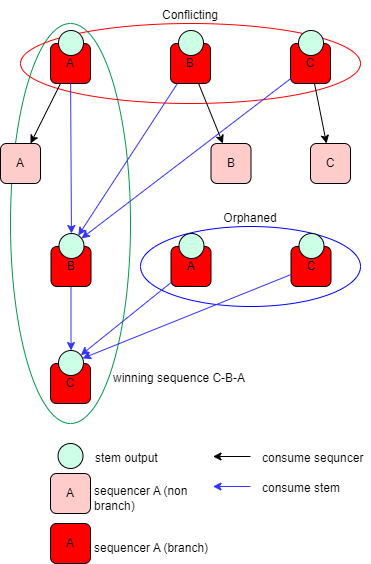}
    \caption{Tree of branches}
    \label{fig:tree}
\end{wrapfigure}

\begin{definition}[\Gls{branch transaction}]
A \gls{sequencer transaction} $T$ that is on the slot edge (with $ticks(T) = 0$) is known as a branch transaction or \textbf{branch}.    
\end{definition}

Branch transactions cannot endorse other sequencer transactions, because cross-slot \glspl{endorsement} are invalid.

We single out branches, as a special form of \glspl{sequencer transaction}, to serve \textbf{several important purposes}.

\subsubsection{Stem output}
\begin{definition}[\Gls{stem}]
The stem output is a special output produced by a branch transaction with a \textit{ stem lock} validation script on it. Only branches produce stem outputs.
\end{definition}

Each branch transaction, in addition to being sequencer transaction, is required:
\begin{itemize}
    \item to produce exactly one stem output;
    \item to consume exactly one stem output from some other branch, called \textbf{stem predecessor};
\end{itemize}    

\begin{definition}[Stem predecessor]
    The \textit{stem predecessor} of a branch transaction T is another branch transaction that produced the stem output subsequently consumed by $T$. This relation is denoted as $stemPredecessor(T)$ 
\end{definition}

Note that $stemPredecessor(.)$ makes a relation on the DAG only between branch transactions. It is different from other relations, e.g. predecessors in chains. 

The important trait of the \textit{stem lock} is that it always has 0 amount of tokens on it and any branch transaction can consume it: There is no private key behind the stem lock.

\subsubsection{Purpose of branches: committed ledger states}
The primary purpose of \glspl{branch transaction} is to provide commitment points of the \gls{ledger state} (as discussed in Section \ref{sec:nodes}.\nameref{sec:nodes}).

Each branch transaction serves as a checkpoint with the \textit{Merkle root} for \gls{ledger state} committed to the node database. It establishes a clear point of reference for the committed ledger state. The \textit{Merkle root} is a \textit{cryptographic commitment} representing an immutable ledger state represented by the branch. 

Most of the ledger states eventually become orphaned along with the branches. This creates a need for cleanup and pruning processes to manage the size and complexity of the multi-ledger database effectively (we skip technical details here).

\subsubsection{Purpose of branches: enforce baseline state for each sequencer transaction}
\label{sec:deterministic-baseline}

Each \gls{sequencer transaction} $T$ is enforced to directly or indirectly choose a particular branch transaction $T_{branch}$ in its \gls{past cone} in the same slot, as its baseline. We will elaborate in Section \ref{sec:stem}.\nameref{sec:stem}).

\subsubsection{Purpose of branches: enforce tree of committed ledger states}

\begin{definition}[Tree of branches]
    A sub-DAG formed by branch transactions in the UTXO tangle with $stemProdecessor(.)$ relation as edges is called a \textbf{tree of branches}. 
\end{definition}

By design, all branches on the same edge of the slot are \glspl{conflict}. It is enforced by the fact, that \textit{stem outputs} can be consumed by many other branches as stem predecessor, thus making them \glspl{double-spend}.

This design ensures that only one \gls{branch transaction} and one \gls{stem} can be recorded in \gls{ledger state}. This, in turn, forces the sequencers to select one (preferred) branch among several competing ones to continue to the next slot. 

\subsubsection{Purpose: branch inflation bonus}
A \gls{branch transaction} rewards \gls{sequencer} with a \gls{branch inflation bonus}. Thus, sequencers are incentivized to issue branches. 

Sequencers \textbf{compete} (participate in a cryptographic lottery) to include their branch in the future chain of ledger states to receive this bonus. The \gls{branch inflation bonus} is an enforced capped pseudo-random value that sequencers cannot mine. See also Section \ref{sec:bias}.

%==========================================================================
\subsection{Sequencer baseline} 
\label{sec:stem}
It is required that each \gls{sequencer transaction} had a deterministically defined \gls{baseline state} (see \ref{sec:deterministic-baseline}). Exactly one of the following conditions must be met for a valid sequencer transaction $T$:
\begin{enumerate}[(a)]
    \item $T$ is itself a \gls{branch transaction} and, therefore, defines its own \gls{baseline state}.
    \item The chain predecessor $T_{pred}$ of $T$ ($T_{pred} \leftarrow T$) is in the same slot as $T$. $T$ will inherit its baseline state from $T_{pred}$ recursively.
    \item $T$ endorses (see \gls{endorsement}) another transaction $T'$ on the same slot as $T$: $T'\Leftarrow T$. $T$ will inherit the baseline state from $T'$ recursively.
\end{enumerate}

The enforced validity constraints outlined above ensure that for each sequencer transaction $T$, there is a single \gls{branch transaction}, known as \textit{baseline branch}, on the same slot which represents the \gls{baseline state} of $T$ with the latest \gls{timestamp}. 

The baseline branch is denoted as $baseline(T)$, and the baseline state of $T$ is denoted as $G_{L_T}$. 

For each sequencer transaction $T$ the following is always true: $stem(T) \in S_T$ and $stem(T) \in out(baseline(T))$.

For each branch transaction $T$ also the following is always true: $stem(T) \in out(stemPredecessor(T))$.

The result is that each non-branch \gls{sequencer transaction} $T$ has a chain of \glspl{branch transaction} in its \gls{past cone} along the $stemPredecessor(.)$ relation. 

The chain of branches represents the history of ledger states since the genesis ledger state, which ends with the baseline of the transaction\footnote{chain} of branches, is similar to the chain of blocks in a blockchain. 

To conclude the above:

\begin{statement}
It is guaranteed that each sequencer transaction $T$ in the \gls{ledger} $L$ will always have uniquely and deterministically defined:   
\begin{itemize}
    \item baseline branch transaction $baseline(T)$
    \item stem output $stem(T) = stem(baseline(T)) = stem(G_{L_T})$
    \item baseline state $G_{L_T}$, accessible in the database via its Merkle root
    \item a chain of branches in the past cone, representing history of the baseline state 
\end{itemize}    
\end{statement}

Note that for non-sequencer transactions the statement above does not hold. 

\subsection{Sequencer chain building}

\begin{figure}
\begin{minipage}{0.55\linewidth}
    \centering
    \includegraphics[width=1\linewidth]{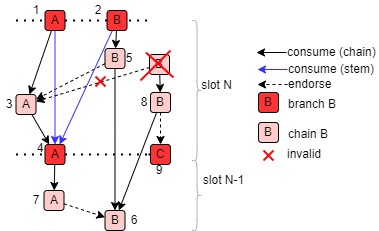}
    \caption{Sequencer chains and branches}
    \label{fig:branches}
\end{minipage}
\begin{minipage}{0.45\linewidth}
    \centering
    \includegraphics[width=1\linewidth]{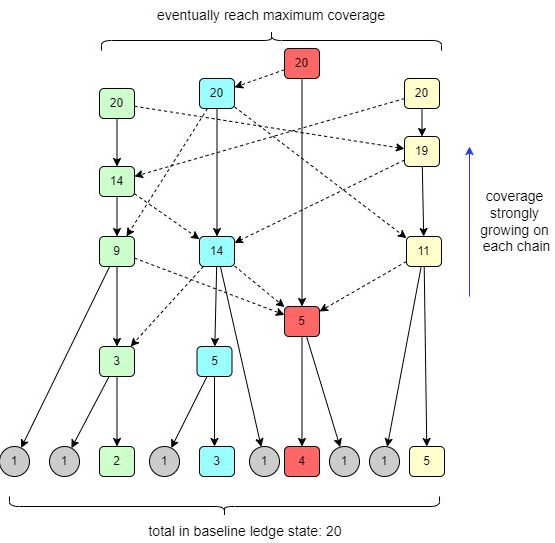}
    \caption{Increasing coverage}
    \label{fig:cross}
\end{minipage}

\end{figure}

In the figure \ref{fig:branches} you can see various ways in which a sequencer can build its chain. 

There are two sequencer chains: chain $A$ and chain $B$. In slot $N-1$, branch $A$ extends into two conflicting branches in slot $N$: Branch $A$ and Branch $B$. This is achieved by both chains consuming the same stem output, resulting in conflicting transactions (transactions 1 and 2) in slot N.

Transaction 3 on chain $A$ extends its own chain from branch 4 within the same slot. Since it continues from branch 4, it inherits the baseline state, so it does not need to endorse any other transaction to reference the baseline.

Transaction 5 continues from its own milestone 6 on the previous slot, so it does not have a baseline defined in slot N. Therefore, it chooses to endorse milestone 3 in another chain. This action defines the baseline of Transaction 5 as branch 4.

The sequencer chain $B$ also continued by creating another milestone 8, following the same predecessor 6 from the previous slot. However, milestone 8 endorsed a different branch 9, which resulted in the two milestones of the sequencer chain $B$ in the N slot conflicting: i.e., 5 and 6 are conflicting.

%=================================================================================
\subsection{Cross-endorsement. Reverting the state}
\label{sec:cross-endorse}

The important implication of sequencer chain building is that any sequencer chain can create its next milestone, by endorsing any other sequencer transaction on another chain  by gradually increasing \gls{ledger coverage} (see \ref{sec:coverage}.\nameref{sec:coverage})

In the conflict-less situation between sequencer chains, sequencers will simply advance by cross-endorsing each other, as in figure \ref{fig:cross} where numbers means \gls{ledger coverage}. However, in a distributed system, forks and conflicting choices made without full information, also due to intentional double spending, are the norm. In that case endorsement by one sequencer transaction of some particular transaction of another sequencer may not be possible.

If the chosen endorsement target conflicts with the latest milestone of the sequencer chain, in order to advance, the sequencer must revert back from its latest state on the chain to a previous state. This requires the sequencer to essentially undo some of its chain progress to reach a state where it can appropriately endorse the chosen target.

The ability to endorse any other sequencer chain with greater \gls{ledger coverage} (see also \ref{sec:coverage}.\nameref{sec:coverage} and \ref{sec:convergence}.\nameref{sec:convergence}) means that a sequencer can continuously increase its ledger coverage by consolidating its state with the states of other sequencer chains. This process of consolidation allows the sequencer to continuously expand its reach and influence over the ledger, providing greater overall contribution to the consensus within the network.

In the scenario described in figure \ref{fig:branches} in the previous section, the sequencer $B$ has just produced milestone 8 but now wants to continue by endorsing milestone 3. However, because milestone 8 conflicts with milestone 3, the sequencer $B$ cannot continue from milestone 8.

Instead, the sequencer $B$ must return (revert) to an earlier state that is consistent with milestone 3. It does this by finding its own chain output (milestone) with \textit{chain ID} $B$ in the baseline state of transaction 3 (the branch 4). Since there is a unique and consistent milestone with \textit{chain ID} $B$ in any state, sequencer $B$ identifies milestone 6 as the own milestone in its past chain that is consistent with the baseline of milestone 3.

Milestone 6 is consistent with milestone 3 because it is within the past cone of branch 4. By reverting to milestone 6, sequencer $B$ can create milestone 5 from milestone 6 and endorse milestone 3. This process is effectively equivalent to reverting the state from milestone 8 to milestone 6 and then continuing from milestone 6.

As a result, milestones 5 and 8 will represent different and conflicting state forks of the sequencer $B$. This situation illustrates how sequencers can revert to earlier states to reconcile conflicts and make consistent progress within the ledger.

It is important to note that the sequencer $B$ does not require the entire transaction 6 to produce transaction 5; it only needs the sequencer output, which is still part of the state. Meanwhile, transaction 6 itself can already be pruned from the UTXO tangle. 

In summary, the example above illustrates how sequencers can revert to a previous state to create a new sequencer transaction that endorses another sequencer transaction on a different chain. Reverting to a prior state, sequencers can resolve conflicts and continue their chains in a consistent manner, ensuring that their milestones remain connected and aligned with the overall ledger state. 

%=================================================================================
\subsection{Ledger coverage}
\label{sec:coverage}
The \gls{ledger coverage} of the sequencer milestone $T$ is a function of \gls{transaction} itself and its \gls{past cone}. Each participant in the distributed system determines it deterministically.

To define the value of \gls{ledger coverage}, we first define a partial value called \textbf{ledger coverage delta} (or simply \textbf{coverage delta}) of the sequencer transaction $T$ in the ledger $L_T$ with the baseline state $G_{L_T}$. The \textit{coverage delta} is the total amount of outputs that are directly and indirectly consumed by the transaction $T$ in its \gls{baseline state} $G_{L_T}$. 

We will denote it by $coverage\Delta(T)$. The past cone of the sequencer transaction $T$ "ends" at \gls{baseline state} $G_{L_T}$. See also Figure \ref{fig:coverage}.

\begin{definition}[Rooted outputs of the sequencer transaction]
Let us assume $seq(T)$ is a sequencer output produced by the $baseline(T)$. Then rooted outputs of the transaction $T$ is a set:
$$
rooted(T) = \{O \in G_{L_T} \| \exists T' \in L_T \text{ where } O \in in(T')\} \cup \{seq(T)\}
$$    
\end{definition}

The output $O \in rooted(T)$ is said to be \textit{rooted}.

The sequencer output produced by $baseline(T)$ may or may not be consumed in the past cone of $T$. However, it is included in the set $rooted(T)$ by definition.

The rooted outputs (except maybe the sequencer output produced by $baseline(T)$) are those that were removed from the baseline ledger state $G_{L_T}$ by the transactions of $L_T$; therefore, they are absent in the final ledger state $S_{L_T}$. $rooted(T)$ is the subset of the baseline ledger state that is \textbf{covered} by the past cone of $T$.

\begin{definition}[Ledger coverage delta of a sequencer transaction]
Ledger coverage delta (or simply coverage delta) of transaction $T$ is the sum of the amounts of the rooted outputs:
$$
coverage\Delta(T) = \sum_{O \in rooted(T)}amount(O)
$$    
\end{definition}

Some observations:
\begin{itemize}
    \item amount of sequencer output produced by $baseline(T)$ is always included into the $coverage\Delta(T)$. So, the minimal value of $coverage\Delta(T)$ is the amount of the sequencer output $seq(T)$ produced by $baseline(T)$;
    \item if $T \leftarrow T'$ then $coverage\Delta(T) \le coverage\Delta(T')$: monotonicity of ledger coverage along consumption references;
    \item if $T \Leftarrow T'$ then $coverage\Delta(T) \le coverage\Delta(T')$: monotonicity of ledger coverage along \gls{endorsement} references. Note, that endorsement references are only possible within one slot;
    \item let's say token holder controls baseline output $O$ on the ledger. It can consume this single output in the new transaction $T$, so it's coverage will be $coverage\Delta(T) = amount(O)$. If the token holder chooses to endorse any other transaction $T'$, $coverage\Delta(T) = amount(O) + coverage\Delta(T') > amount(O)$. So, by consolidating own output with some independent transaction in one state, the token holder will increase ledger coverage of the new transaction;
    \item the coverage delta is capped by the $totalSupply$ on the baseline state: $coverage\Delta(T) \le totalSupply$. So, it cannot grow forever on a given baseline state;
\end{itemize}

We define full \textit{ledger coverage} as a weighted sum of the coverage deltas of the sequence of past branches, with weights that decrease exponentially as they go further back in the sequence. \textit{Ledger coverage} is undefined for non-sequencer transactions.

\begin{definition}[\Gls{ledger coverage} of a \gls{sequencer transaction}]
The ledger coverage (or simply coverage) of the transaction $T$ we define in the recursive form:
$$
coverage(T) = 
\begin{cases}
    initialSupply & slot(T)=1 \\
    coverage\Delta(T) + \frac{coverage\Delta(baseline(T))}{2^{slot(T)-slot(baseline(T))+1}} & \text{T is not a branch}  \\
    coverage\Delta(T) + \frac{coverage\Delta(stemPredecessor(T))}{2^{slot(T)-slot(stemPredecessor(T))}} & \text{T is a branch} \\
\end{cases}
$$ 
\end{definition}

Let us say $T=B_N$ is a branch transaction and $B_N, B_{N-1}, \dots B_0$ is a chain of branches back to genesis, $B_{i-1}=stemPredecessor(B_i), i= 1\dots N$. (note that some slots may not contain a branch). Then the above recursive definition for branch transaction $T$ translates into to following:

$$
coverage(T) = 2 \cdot \sum_{i=1}^{N}\frac{coverage\Delta(B_i)}{2^{slot(B_i)-slot(B_{i-1})}}
$$

For practical reasons, we use integer arithmetic and assume that the values of $coverage$ fit 64-bit integers; therefore, we can restrict summing along the maximum $N=64$ slots back.

$coverage\Delta(T) \le totalSupply$. In means 
$$
coverage(T) \le totalSupply \cdot (1 + \frac{1}{2} + \dots + \frac{1}{2^{63}}) < 2 \cdot totalSupply
$$

Note that in the same manner as above, we can define $coverage\Delta(\cdot)$ and $coverage(\cdot)$ for the set of transactions with the consistent past and the same baseline, by taking the union of $rooted(\cdot)$. In that case, we will use notation $coverage\Delta(\{T_1, \dots T_K\})$ and $coverage(\{T_1, \dots T_K\})$ respectively.

%==========================================================================
\subsection{Sequencing strategy}
\label{sec:strategy}

We introduced the general idea in the section \ref{sec:cooperative}. Here, we provide a more detailed discussion. 

The sequencing strategy is a complex, implementation-intensive, and heuristic-based problem. Although we strive to present the approach in a generic manner which is based on the implementation of testnet, we acknowledge the lack of a deeper mathematical model and a generic framework for the sequencer strategy at this time. The testnet implementation demonstrates the feasibility of practical sequencer strategies, and we anticipate ongoing research and continuous improvement of existing strategies.

It is important to note that the sequencer, just as any token holder, is free to choose any strategy it likes, as long as it conforms to the ledger constraints. The Proxima can be seen as a for-profit game among token holders with rules defined in the form of ledger constraints. In theory, multiple strategies are possible, each with pros and cons. According to our claim, all of them are based on the \gls{biggest ledger coverage} rule. 

The \textbf{sequencing strategy} refers to the algorithm or behavioral pattern used by a program (or, potentially, an AI agent) to guide the actions and decisions of a sequencer within the network.
In the provided algorithm (\nameref{alg:sequencer}), a generic form of one possible strategy is described: \mbox{\Call{GenericSequencer}{.}}.

Our hypothesis is that more efficient strategies will have only marginal advantage against the generic one in terms of the profitability, or, if not, they will be immediately copied by other sequencers. 

Technically, \gls{sequencer} is an automated process that typically has access to the private key that controls the token holder's tokens. It also interacts with the UTXO tangle and other services provided by the node. In practice, a sequencer is a program that runs as (an optional) part of the node. 

A sequencer must build a sequencer chain in the ledger by maximizing interests of the token holder, aiming for profit and return on capital, based on income from inflation, tag-along fees and delegation margin (see also \nameref{sec:non-sequencer}). 

The sequencer is constantly monitoring the local \gls{UTXO tangle}, which is dynamically updated with new transactions as they arrive, and continuously producing sequencer transactions in real time to build on the sequencer chain. 

There are two primary tasks that define the sequencer's strategy: $\Call{decideTimestampTarget}{\cdot}$ and $\Call{selectInputs}{\cdot}$ (see descriptions below).

The task $\Call{selectInputs}{\cdot}$ can be computationally expensive in high-load situations (high \textit{transactions-per-second}).

These tasks may impose significant real-time constraints on the sequencer, affecting its overall performance.

\begin{algorithm}[H]
\caption{Generic sequencer}\label{alg:sequencer}
\renewcommand{\algorithmicrequire}{\textbf{Input:}}
\begin{algorithmic}[1]
\Procedure{GenericSequencer}{$sequencerID, privateKey, utxoTangle, localClock$}
\State $timestampTarget \gets 0$
\Loop
    \State $bestCandidate \gets nil$ \Comment{assume $coverage(nil)=0$}
    \State $timestampTarget \gets \Call{decideTimestampTarget}{localClock.now, timestampTarget)}$
    \While{$localClock.now < timestampTarget.realTimeValue$}
        \State $chainTip, endorse, consume \gets \Call{SelectInputs}{sequencerID, timestampTarget, utxoTangle}$
        \State $candidate \gets \Call{MakeTransaction}{privateKey, timestampTarget, chainTip, endorse, consume}$
            \If{$coverage(candidate) > coverage(bestCandidate)$}     \State $bestCandidate \gets candidate$
            \EndIf
    \EndWhile
    \If{$bestCandidate \ne nil$}
        \State \Call{submitTransaction}{bestCandidate}
    \EndIf
\EndLoop    
\EndProcedure

\end{algorithmic}
\end{algorithm}

\subsubsection{Pace of the sequencer chain}
Function $\Call{decideTimestampTarget}{\cdot}$ is responsible for calculating the timestamp of the next sequencer transaction, effectively setting the pace of the chain. The optimal pace, or frequency of sequencer transactions, is a matter of strategy and optimization, and the sequencer must consider the following factors.

\begin{enumerate}[(a)]
    \item Transactions are subject to a minimum interval between timestamps, defined by the \textbf{transaction pace validity constraint}. This constraint sets a lower bound on the timestamp for the next sequencer transaction based on the given inputs. One of objectives behind pace constraints, is to put a cap on maximum transaction issuance rate per token holder (see \nameref{sec:anti-spam}).
    
    \item The sequencer must produce the next sequencer transaction in real time without issuing transactions too frequently or with excessive gaps between them. Producing transactions too frequently may hinder consolidation of ledger coverage and disrupt synchronization with other sequencers. Conversely, long gaps between transactions may limit the sequencer's ability to process tag-along and delegation outputs and disrupt other sequencers' ability to catch up.
    
    \item The target timestamp cannot be too far ahead or behind real time (see \nameref{sec:nodes}) to avoid that sequencer transactions are placed on hold (if ahead of time) or ignored (if perceived being late).
    
    \item The sequencer aims to generate a branch transaction on the edge of the slot with the maximum possible coverage.
\end{enumerate}
Taking into account the above factors, choosing an optimal next-timestamp target typically relies on heuristics. Fortunately, experiments on the testnet have shown that reasonable heuristics exist, providing a foundation for optimization.

\subsubsection{Select inputs for the transaction}
When the $ticks(timestampTarget)=0$, the sequencer is simply creating a branch transaction. There is little to optimize in this scenario, as the only inputs needed for the branch transaction are the chain predecessor and the \gls{stem} of the predecessor's \gls{baseline state}.

In other cases, the function $\Call{selectInputs}{\cdot}$ analyzes the UTXO tangle and returns a collection of consistent inputs needed to construct the transaction optimally. There can be a large number of such collections, and the function is stateful, keeping track of all the collections it has previously returned. Subsequent calls to $\Call{selectInputs}{\cdot}$ enumerate all or a reasonable majority of all possible collections and return them in heuristic order, prioritizing those that provide the most significant \gls{ledger coverage} first.

The function $\Call{selectInputs}{\cdot}$ must choose the following inputs from the UTXO tangle for each candidate transaction:
\begin{itemize}
    \item This output will be used as a predecessor in the transaction and will form the basis for the chain.
    \item The function should identify and select a set of endorsement targets, which are transactions that maximize the ledger coverage of the future transaction. This selection should prioritize maximizing coverage to the greatest extent possible.
    \item Tag-along outputs provide income for the sequencer and increase ledger coverage. The function should strive to maximize the consumption of tag-along and delegation outputs to enhance the sequencer's income and coverage.
\end{itemize}
The function $\Call{selectInputs}{\cdot}$ must ensure that the entire set of selected inputs does not contain \glspl{double-spend} in the consolidated \gls{past cone}. As \gls{conflict} detection has high algorithmic complexity in \gls{transaction DAG}, it is a challenge to minimize latencies when analyzing the UTXO tangle.

Suppose that at some point in time, the UTXO tangle contains $N$ \glspl{sequencer milestone}, $M$ \glspl{tag-along output}, $K$ \gls{delegation} outputs that can be potentially consumed by the sequencer. In addition, there are $E$ potential \gls{endorsement} targets, out of which $P$ we want to endorse in the transaction ($P$ usually is 1 or several). There are a vast number of possible combinations for the collection, candidates for selection: $O(N\cdot2^M\cdot2^K\cdot \binom{E}{P})$. 

For each of these combinations, the function will have to check if the combination does not contain \glspl{double-spend} in the \gls{past cone}. In this way the task of selecting the optimal combination may become too computationally costly for real-time optimization. This points to the importance of the heuristics. 

To avoid complexities, we will skip details of $\Call{selectInputs}{\cdot}$ as it is implemented in the testnet version. 

In general: $\Call{selectInputs}{\cdot}$ first finds the heaviest other sequencer transaction in the slot and iterates over its own past milestones to find the best one to be combined with the heaviest milestone. At least one own milestone consistent with the endorsement target always exists, but it may not be the heaviest one, so the sequencer needs to try a set of possibilities to select the best consistent extent/endorse pairs.

Then, for each potential extent/endorse combination, the sequencer collects as much pending tag-along and delegation outputs as it fits into the transaction. Each of those combinations is verified to have a conflict-free past cone. The consistent collection is returned.

Note that repetitive call to $\Call{selectInputs}{\cdot}$ will return different results because the situation on the UTXO tangle is constantly evolving.

This simple strategy assumes only one endorsement target. It results in a minimal rate of state consolidation and coverage growth with each transaction. Experiments show that even with this limitation, the convergence of the state toward consensus is relatively fast. Strategies that consolidate two or more sequencer chains, as well as strategies that are more adaptive to the current situation on the network and, therefore, significantly increase the rate of convergence (see also \nameref{sec:convergence}).  

%==========================================================================
\subsection{Non-sequencing users}
\label{sec:non-sequencer}
\noindent
The sequencers will be the primary proactive participants in the network. Network security depends on the cooperation and agreement of the main players. However, what about other users who cannot or do not want to be sequencers for various reasons?

Two types of mechanisms—\textit{tagging-along} and \textit{delegating}—allow non-sequencer users to contribute to the consensus on the ledger state and earn inflation rewards.

Each output in the ledger must include a mandatory \textit{lock constraint} that enforces certain requirements for the consuming transaction, such as unlocking conditions. A common example is the \textit{addressED25519} lock, where the output can be unlocked with a transaction signature corresponding to a specific address, or a \textit{chain lock}, where the output can be unlocked with the signature on a transaction that unlocks a chain output with a specified \textit{chain ID} in the same transaction, that is, the chain controller's signature.

%==========================================================================
\subsubsection{Tagging-along. Feeless option}
\label{sec:tag-along}

If \textit{Alice} wants to transfer an amount of tokens to \textit{Bob} and \textit{Alice} is not a \gls{sequencer}, she can create a transaction $T$ that consumes some of her outputs and produces an output of amount $A$ with the target lock constraint set to \textit{addressED25519} using \textit{Bob}'s wallet address. Additionally, $T$ would produce a remainder output that goes back to \textit{Alice}'s wallet.

However, if Alice submits $T$ to the network, the transaction will not be processed because there is no sequencer to include the output in the consensus ledger state. As a result, $T$ will become \gls{orphaned}, i.e., it will be ignored by the network.

To fix that, in the transaction $T$, \textit{Alice} produces an additional output with an amount of tokens $\phi$ and \textit{chain-locks} it to make it unlockable by a sequencer with the specified \textit{sequencerID}. In this way, the transaction $T$ produces three outputs: (1) with amount $A$ to \textit{Bob's} address, (2) with amount $\phi$ to chain $sequencerID$ and (3) the remainder.    

The output (2) with the amount $\phi$ that is chain-locked to the sequencer with \textit{sequencerID} is known as \gls{tag-along output}. The amount $\phi$ is called \gls{tag-along fee}, and the sequencer with \textit{ sequencer ID} is known as the \textbf{target tag-along sequencer}.

When \textit{Alice} creates a transaction $T$ that includes a \gls{tag-along output} targeted at a specific sequencer, the sequencer becomes incentivized to consume that output because it provides income in the form of \gls{tag-along fee}. Consequently, the sequencer naturally includes the tag-along output in its own sequencer transaction.

By incorporating the transaction $T$ into its own transaction, the sequencer ensures that $T$ becomes part of its \gls{ledger state}, which is likely to be consolidated with other ledger states. This allows \textit{Bob} to receive his tokens as intended, and the sequencer earns \gls{tag-along fee}. In this way, \textit{Alice}'s transaction is successfully processed, and both the target recipient (\textit{Bob}) and the sequencer benefit from the transaction.

The amount $\phi$ of the tag-along fee is generally small compared to the amount being transferred to \textit{Bob} ($A$), which can vary in size. The amount of the tag-along fee is determined by the sequencer and, ultimately, by market forces.

\textit{Alice} is naturally motivated to choose tag-along sequencers that offer smaller fees, as this will minimize her transaction costs and make her transaction more cost-effective. This market-driven competition among sequencers to provide lower fees encourages efficiency and cost effectiveness in the network, benefiting users like \textit{Alice}. 

If Alice is running her own sequencer, she will obviously choose to tag her transaction along her own sequencer and pay the fee to herself, thus making the transaction \textit{feeless}. 

\textit{Alice} contributes to the ledger coverage of the sequencer transaction with the full amount of the transaction, not just the tag-along fee $\phi$. This contribution to coverage can include the amount transferred to \textit{Bob} ($A$) and any remaining funds returned to Alice as change.

Sequencers have an inherent interest in maximizing ledger coverage to improve their chain's reach and influence within the network. This dynamic creates the potential for larger transfers to be entirely \textbf{feeless}. Since sequencers have an independent interest in attracting more coverage to their chains, they may choose to process larger transactions without imposing a fee to gain the associated ledger coverage benefits. 

The tag-along mechanism allows non-sequencer users to earn inflation just like sequencers. \textit{Alice} produces transaction with the chain transition and inflation on it. She tags-along the transaction to some sequencer. If the tag-along fee is less than the inflated amount, the transaction will be profitable for \textit{Alice}. 

Leaving tag-along output with zero or very small fees creates also a problem of \textit{dusting}\footnote{huge number of small amount UTXOs which bloat the database}. We will elaborate on the solution for that problem in section \ref{sec:anti-spam}.\nameref{sec:anti-spam}.

%==========================================================================
\subsubsection{Delegation}
\label{sec:delegate}

\textbf{Delegation} allows passive capital to be involved in cooperative consensus without the token holder actively participating in sequencing. This is achieved by allowing the token holder to delegate the rights to use their capital as ledger coverage to a sequencer while maintaining ownership and control over their tokens.

In addition, delegation solves the scalability bottlenecks inherent to massive participation of the token holder in the consensus. See chapter \ref{sec:scalability-delegation}.\nameref{sec:scalability-delegation} for a more detailed description.

%==========================================================================
\section{Convergence} 
\label{sec:convergence}

\subsection{Growing the UTXO tangle}

This section explains how sequencers can build \gls{UTXO tangle} continuously, forever.

For clarity, we consider a simplified scenario where tagging-along and delegation are absent, and every existing sequencer is present in any relevant branch ledger state. However, the following reasoning remains valid even without these simplifications. 

\begin{statement}
Every \gls{sequencer} can create a \gls{sequencer transaction} in the slot, which starts with at least one branch transaction. 
\end{statement}
\begin{proof}
Let us say that we have $N$ branches that start the slot: $B_1, \dots B_N$. For a sequencer with ID $A$, there are two (non-exclusive) possibilities:
\begin{enumerate}[(a)]
    \item one of transactions $B_i$ belongs to the sequencer $A$. In this case, the sequencer just creates the successor $T_{next}$ of the transaction $B_i$ in the same slot by consuming its \gls{sequencer milestone} of $B_i$ as the predecessor. The baseline of the transaction will be the branch itself: $baseline(T_{next})=B_i$ 
    
    \item sequencer $A$ has its tip in at least one of branches ledger states, say $B_i$: $tip_{B_i}(A)$. The tip is guaranteed to be unique. The sequencer $A$ can always create successor $T_{next}$ by consuming $tip_{B_i}(A)$ and endorsing $B_i$. The latter becomes the baseline of $T_{next}$
\end{enumerate}
\end{proof}

\begin{definition}[Full sequencer coverage]
We say that \textit{full sequencer coverage is reached in the slot $\tau$ by a transaction $T$} when the set $rooted(T)$ contains all sequencer milestones in the state $baseline(T)$ (in other words, when transaction $T$ covers all sequencers in its baseline). 
\end{definition}

When full sequencer coverage is reached in the slot by some transaction $T$, coverage of any new transaction (assuming no tag-along and delegation inputs) cannot be greater than the reached maximum $coverage\Delta(T)$.

\begin{statement}{Maximal coverage among transactions of the UTXO tangle in one slot can grow as long as full sequencer coverage is not reached in that slot and time constraints permit}
\end{statement}
\begin{proof}
Let us say $T$ is the transaction with the biggest coverage in the slot. By assumption, some sequencer $A$ is not covered by $T$, that is, $tip_{baseline(T)}(A) \notin rooted(T)$.

The sequencer $A$ can always create transaction $T_{next}$ by consuming its own milestone $tip_{baseline(T)}(A)$ and endorsing $T$. The coverage of the new transaction will be:
$$
coverage\Delta(T_{next}) = coverage\Delta(T) + amount(tip_{baseline(T)}(A)) > coverage\Delta(T)
$$
\end{proof}

\begin{statement}{Sequencers can build UTXO tangle continuously}
\end{statement}
\begin{proof}
We start with some \gls{slot} $\tau$. According to the statements above, each sequencer can create transactions in the slot. After adding a new transaction to the slot, one of transactions will have maximal coverage. Until full sequencer coverage is reached and ledger time constraints permit, new transactions can keep increasing ledger coverage.

After reaching full sequencer coverage or the end of the slot $\tau$, sequencers can create branches at the edge of the next slot $\tau+1$. 

This process can be repeated again and again. 
\end{proof}

%==========================================================================
\subsection{Rich chain bias. Meta-stability. Random branch inflation}
\label{sec:bias}

Each \gls{sequencer} is trying to reach as large \gls{ledger coverage} in the slot as possible. This leads to the conclusion that it is highly likely that the ledger coverages of branches on the next slot will be close to each other or even equal. It can lead to a situation of meta-stability. 

Meta-stability in this context refers to a state where the system is stuck or oscillates between different branches without a clear preference for any single one due to their equal coverages. While meta-stability may be broken by other random factors, in theory it could lead to slower convergence or even in parallel chains of branches (forks) with approximately equal coverage. This may create a problem, known as \textbf{nothing-at-stake}, when sequencers, by blindly following \gls{biggest ledger coverage}, same time maintain several conflicting chains of branches (forks) in parallel.

The naive approach to address this problem would be, in case of equal coverage, to prefer higher transaction hash. However, this leads to an undesirable hash mining competition among sequencers.

\begin{wrapfigure}{R}{0.55\textwidth}
    \centering
    \includegraphics[scale=0.6]{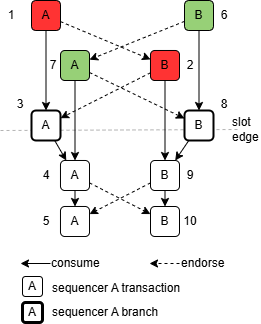}
    \caption{Preferring greater branch inflation}
    \label{fig:bias}
\end{wrapfigure}

Let us consider the figure \ref{fig:bias} ($T_i$ will designate the transaction with the number $i$ in the picture). Let us say that \glspl{sequencer milestone} of the sequencers $A$ ($T_4$) and $B$ ($T_9$) at the end of the previous slot have respective token amounts $amount_A$ and $amount_B$. 
Furthermore, let us assume that the \glspl{ledger coverage} of the transactions ($T_4$) and ($T_9$) are equal. This means that the coverages of branches $A$ and $B$ are also equal: 
$$ 
coverage (T_3) = coverage (T_8) = C
$$

The amounts in \glspl{sequencer milestone} of the branches $A$ ($T_3$) and $B$ ($T_8$) are $amount_A + I_A$ and $amount_B + I_B$, where $I_A$ and $I_B$ are \gls{branch inflation bonus} amounts.

Now let us consider two options, labeled as \textbf{red} and \textbf{green}, for how the sequencer $A$ could continue with a sequencer transaction to the next slot:
\begin{itemize}
    \item The red option means that $A$ will consume its own branch as a predecessor and continue with the red transaction $T_1$ that endorses $T_2$ as successor. The coverage of it will be: $coverage(T_1) = amount_A + I_A +amount_B + \frac{C}{2}$.
    
    \item The green option means that $A$ will endorse the branch of $B$, $T_8$, and will consume its own output as the predecessor $T_4$ from the baseline state of $B$, $baseline(T_8)$. The coverage of the transaction $T_6$ will be $coverage(T_7) = amount_A + amount_B + I_B + \frac{C}{2}$
\end{itemize}

Between the two possible options $T_1$ and $T_7$ the sequencer $A$ and all subsequent sequencers will choose the one with the largest coverage, which, in this case, is the one with the greater \gls{branch inflation bonus}. Symmetrically, the same situation will occur with the sequencer $B$. 

In case of $I_A > I_B$, the red option will win while the green option will never appear or will be \gls{orphaned}. In case of $I_A < I_B$, the green option will prevail.

\begin{conclusion}
In the frequent situation where \glspl{ledger coverage} are equal between branches, the branch with the larger \gls{branch inflation bonus} will prevail, i.e. the \gls{biggest ledger coverage} rule, followed by sequencers, will point to the one with bigger branch inflation bonus. The amount of tokens on the \glspl{sequencer milestone} does not influence the outcome.
\end{conclusion}

If $I_A = I_B$, the situation will be indecisive, that is, may lead to a meta-stability. We conclude that fixed (as ledger constant) branch inflation bonus would not be a good idea because of this reason.

On the other hand, if we assume the branch inflation bonus is proportional to the on-chain balance (the same way inflation is calculated inside the slot), it would result in a situation where the richer chain will always win the branch inflation bonus, i.e., what we call \textit{rich chain bias}. While this may not pose a security problem, though it would be an unfair reward distribution which would disincentivize smaller sequencers from issuing branches.

We introduce an enforced pseudo random \gls{branch inflation bonus} to address both problems: meta-stability \textbf{and} rich chain bias. The random branch inflation bonus removes any advantage for the richer sequencer, and it also breaks meta-stability.  

The \gls{biggest ledger coverage} rule becomes two-fold:
\begin{enumerate}
    \item If coverages are different, the transaction with the bigger one prevails.
    \item If coverages are equal (rare case), the transaction $T$ with the largest transaction hash $id(T)$ prevails.
\end{enumerate}

We also want to make any transaction hash mining race impractical by making Option 2 above (collision of two coverages equal) a highly unlikely event.

We adopt the approach of \textit{Verifiable Random Function} (VRF)\footnote{\url{https://medium.com/algorand/algorand-releases-first-open-source-code-of-verifiable-random-function-93c2960abd61}}:  

\begin{itemize}
    \item Each sequencer derives its \textit{secret key} $SK$, used in VRF, from its private key, used for transaction signatures. For simplicity, here we assume that \textit{verification key} $VK$, used in VRF functions is the same public key \textit{ED25519} used in signatures and therefore always available on transaction (if this is not technically true, it does not change the principle).
    \item with each branch transaction, sequencer computes two values:
    $$
    Evaluate(SK, s) \rightarrow (rnd, proof)
    $$
    where $s = VRF_{pred}||slot$ is the \textbf{value of VRF of the predecessor branch transaction concatenated with the slot number}, and $rnd$ is generated random value, $proof$ is the cryptographic proof of randomness. ($s$ is enforced and never repeats)
    \item The value $I = rnd \mod (I_{max}+1)$ is used as \gls{branch inflation bonus}.    
    \item The following constraint on values $proof$ and $rnd$ is enforced on the produced \gls{sequencer milestone}:
$$
Verify(VK, s, rnd, proof) = true \And I = rnd \mod (I_{max}+1)
$$
\end{itemize}

The constant $I_{max}$ is a ledger constant of the maximum possible branch inflation bonus.

In this way, each branch will have an enforced pseudo-random value of the inflation bonus in the interval $0<I\le I_{max}$, which is derived by each sequencer for its branch transaction from its private key.

The properties of the VRF guarantee that the sequencer cannot produce different values of $rnd$ for the same $VK$ (the public key) by manipulation. Chances that two equal inflation bonus collide in the same slot is one in $I_{max}^2$\footnote{$\approx 10^{-14}$ in the testnet}. This makes any attempt to gain advantage by mining the transaction ID impractical.  

\subsection{Randomness of convergence. Heuristics}

The growth of the UTXO tangle is a random process not only because of the random \gls{branch inflation bonus}, but also due to other unpredictable variables, such as communication delays, the random time needed to produce the transaction and sequencer strategy. Each sequencer selects inputs for the next transaction in the chain from an asynchronously changing set of transactions that arrive at random intervals and in random order from other sequencers. The produced transaction reaches other sequencers after a random time interval.

The mathematical modeling of this random process is beyond the scope of this paper. Meanwhile, we use our common judgment to come up with some general guidelines for the sequencers.

\begin{wrapfigure}{r}{0.5\textwidth}
    \centering
    \includegraphics[scale=0.45]{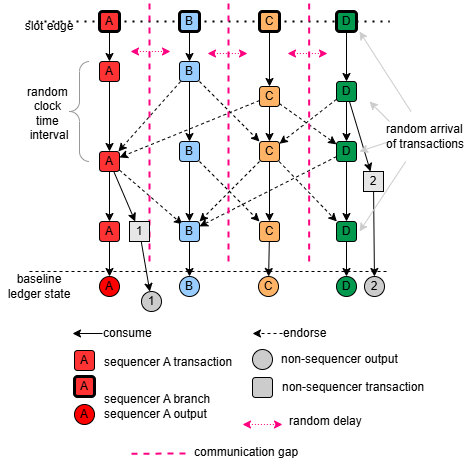}
    \caption{Random converging}
    \label{fig:convergence}
\end{wrapfigure}

Figure \ref{fig:convergence} illustrates a possible situation on the UTXO tangle. For simplicity's sake, the diagram assumes that all sequencer transactions share the same baseline state.

Each sequencer is separated from other sequencers by network communication delays. Sequencers $B$ and $C$ successfully cover the entire \gls{baseline state}, with all outputs except $2$. Thus, their coverage will be equal (largest), causing them to compete as branches. The branches of $B$ and $C$ have good chances to be extended even further.

Meanwhile, the sequencers $A$ and $D$ did not cover all possible outputs in the baseline due to network delays. Their coverage will be smaller, and their branches and respective ledger states will have no chance of being chosen as preferred, so they will eventually be \gls{orphaned}.

The non-sequencer output $1$ was tagged along with sequencer $A$ and then endorsed by sequencers $B$ and $C$, so it will make it into the ledger state in the next slot anyway, even if sequencer $A$ has no chance of producing the preferred branch and regardless of which of the two heaviest branches wins. The output $2$ was tagged along to sequencer $D$ but will not be included in the state in the next slot (although it can be tagged along in the next slot).   

\begin{observation}
\text{  }
\begin{enumerate}
    \item Sequencer transactions with more tokens are more likely to be chosen as endorsement targets by others, leading to faster inclusion in the (future) ledger state. 
    \item Employing delegated capital in the sequencer and consuming tag-along inputs helps the sequencer transaction to be included in the (future) ledger state more quickly. 
    \item Endorsing more other sequencer chains generally leads to faster convergence than endorsing fewer. 
    \item Smaller communication delays between the heaviest sequencers favor convergence.
    \item Submitting a transaction later (in real terms) than its timestamp may result in it being no longer the heaviest by the time it reaches other sequencers. Therefore, a \textbf{local clock that is behind the average is unfavorable for convergence}.
    \item Submitting a transaction too early (in real terms) than its timestamp may result in the nodes putting it on hold, causing the submitting node to lose the opportunity to wait a few extra milliseconds and endorse a heavier transaction. Therefore, a \textbf{local clock that is ahead of the average is unfavorable for convergence}.
    \item The sequencer must consume the tag-along outputs as early as possible, making the sequencer chain heavier, thus boosting convergence.
    \item Many sequencers will likely come to the end of the slot with equal (or almost equal) ledger coverage, which is close to the maximum possible. This will lead to multiple competing branches with equal coverage at the slot edge. 
    \item To be competitive for the branch inflation bonus, sequencers must cooperate in advance.
    \item Lazy, isolated, or non-cooperative sequencers have a smaller chance of winning the branch bonus, if any.
\end{enumerate}
\end{observation}

%==============================================================================================
\section{Security}
\label{security}

Here, we present \textbf{a sketch of the security model}. Deeper modeling, which takes into account more aspects of safety and liveness is a topic of future research.

Unlike the \gls{bft} committee, where participants can deduce the finality of a transaction from voting data collected from other participants, in the system with an unbounded and unknown set of participants, the finality of the transaction is probabilistic. That is, if \textit{Alice} sends transaction to \textit{Bob}, he must \textbf{ assess the chances} of the transaction being reverted. Only if the likelihood to be reverted becomes negligible, \textit{Bob} can release goods or money to \textit{Alice}. 

The finality rule in a probabilistic consensus is defined by \textit{Bob} himself. An example of a subjective finality rule often used in Bitcoin is waiting for a transaction to be buried under at least 6 blocks as per the "six-block rule. Let us define a similar subjective finality rule for cooperative consensus. 

\begin{definition}[Finality rule with parameter $\theta$]
\label{def:finality}
To accept the transaction $T$ as final, \textit{Bob} waits for a branch $B_0$ with finite sequence of branches $B=(B_0, B_1, \dots B_{N-1})$ where $B_{i+1}=stemPredecessor(B_{i})$ for $i=0, \dots N-2$ and $T\in B_i$, $i=0, \dots N-1$ and for which:
$$
\beta_T(B_0, \theta) = \sum_{i=0}^{N-1}\frac{coverage\Delta(B_i)}{2^i} > 2\cdot \theta \cdot totalSupply
$$
\end{definition}
where $1/2 < \theta < 1$ is a parameter assumed by \textit{Bob}.

Note that we can assume that the chain of branches $B_0, \dots B_{N-1}$ is "longest" in the sense that for $B_N=stemPredecesor(B_{N-1})$, $T\notin B_N$. In this case:
$$
coverage(B_0) = \beta_T(B_0,\theta)+\frac{coverage(B_N)}{2^{N}} 
$$
Obviously, $coverage(B_0) - \beta_T(B_0,\theta)=\frac{coverage(B_N)}{2^{N}} <  \frac{totalSupply}{2^{N-1}} \rightarrow 0$, so $\beta_T(B_0,\theta)\approx coverage(B_0)$ for large $N$.

%================================================================================
\subsection{Safety}
\label{sec:safety}

Let us assume \textit{Bob} receives branch $B_0$ that satisfies the finality criterion above and $slot(B_0)=\tau$. We assume that branch timestamps are closely synchronized with real-time reference time, as perceived by Bob's local clock (see \ref{sec:ledger-time} \nameref{sec:ledger-time}). 

The branches with timestamp $\tau$ arrive at \textit{Bob} in a random order. Among the branches that satisfy the finality criterion, we assume $B_0$ exhibits the largest ledger coverage.  

The central question of safety is under what conditions could an alternative branch $B'_0$ arrive within the duration of the slot $\tau$, where $\text{slot}(B'_0) = \tau$ and $\text{coverage}(B_0') > \text{coverage}(B_0)$, but without containing transaction $T$ (i.e., $T \notin B_0'$)? 

% The question is: Under what conditions will another branch $B'_{0}$ with $slot(B'_{\tau})=\tau$ and $coverage(B'_{0}) > coverage(B_{0})$ but with $T \notin B'_{0}$, arrive during (the rest of) the slot $\tau$? 

A potential adversary, \textit{Eve}, could construct $B'_0$ to induce the network to adopt this branch instead, thus excluding \textit{Bob}'s transaction $T$. This attack can be regarded as a variant of a \textbf{long-range attack}.

% A malicious actor \textit{Eve} could produce $B'_{0}$ thus causing the network to switch to branch $B'_{0}$ and \textit{Bob} transaction $T$ will not be in it. We can see this attack as a sort of \textbf{long-range attack}. 

To analyze the feasibility of such an attack, we seek to establish a lower bound on the amount of capital, under \textit{Eve}'s control, required to successfully execute it. We observe that the quantity  

% We want to assess the lower bound of "malicious" capital \textit{Eve} would need to be controlled to make a successful attack. We observe that the amount of coverage

$$
H = 2\cdot totalSupply - coverage(B_{0})
$$ 
represents tokens that did not participate in the past cone of $B_0$. These tokens could potentially be included in $B'_0$ without altering their position in $T$. Such tokens may be considered \textit{uninformed} or \textit{innocent}, as they were excluded from $B_0$ due to communication failures, network partitioning, or similar reasons.

% did not participate in the past cone of $B_{0}$.
% So, these tokens could participate in $B'_{0}$ without changing their "opinion" on $T$. That would be "innocent" or "naive" coverage, that does not participate in $B_{0}$ due to lack of communication, network partitioning, or a similar reason.

% Essentially, the attack would consist of keeping the "innocent" tokens isolated from the reach of the main amount of token holders \textbf{and}, in secret, consolidating a certain amount of "malicious" tokens (which already covered $T$) with the "innocent" ones to leave $T$ outside the coverage.  

The essence of the attack involves isolating these \textit{innocent} tokens from the influence of the broader token-holding majority while covertly consolidating a sufficient fraction of \textit{malicious} tokens—previously contributing to the coverage of $T$ with the \textit{ innocent} tokens. The objective is to ensure that $T$ is excluded from the resulting ledger coverage.  

Let us denote $coverage(B'_{0})=H+M$, where $M$ is "malicious" coverage which includes $T$ in $B_{0}$ but excludes it in $B'_{0}$:
$$
M = coverage(B'_{0}) - H = coverage(B'_{0}) - 2\cdot totalSupply + coverage(B_0) > 2\cdot(coverage(B_0)-totalSupply)
$$
or
$$
M  > 2\cdot(coverage(B_0)-totalSupply) > 2\cdot totalSupply \cdot (2\cdot\theta-1)
$$

So, the amount of coverage $M$ needed to flip the branch $B_0$ grows together with $coverage(B_0)$. 
It may be too small when $coverage(B_0)$ is close to $totalSupply$ (or $\theta$ is close to $1/2$). The requirement can become excessively high if $coverage(B_0)$ is close to the maximum possible value $2\cdot totalSupply$ (or $\theta$ is close to $1$). 

Let us assume $B_N=B'_N$ is the branch which is shared by chains $\{B_i\}$ and $\{B'_i\}$ in their past cone: the "point of fork", because $T\notin B_N$.

Let us define the value $\beta'= coverage(B'_0) - \frac{coverage(B_N)}{2^{N}}$

To satisfy condition $coverage(B'_0)>coverage(B_0)$, we need $\beta' > \beta_T(B_0,\theta)$. 

We can represent $\beta'$ as sum
$$
\beta' = \sum_{i=0}^{N-1}\frac{coverage\Delta(B'_{i})}{2^{slot(B_0)-slot(B'_{i})}}
$$ 
over the chain of branches $B'=(B'_0, B'_1 \dots B'_{N-1})$, where $T\notin B'_{i}$, which is parallel to chain $B_0, \dots$ by assuming $B_i=nil$ (a special symbol of absence) and $coverage\Delta(nil)=0$ for the branches missing in the slot $i$. This way:
$$
\beta'=\sum_{i=0}^{N-1}\frac{coverage\Delta(B'_i)}{2^i} > \sum_{i=0}^{N-1}\frac{coverage\Delta(B_i)}{2^i}=\beta_T(B_0, \theta)
$$

We should assume $coverage\Delta(B'_i) \le coverage\Delta(B_i)$ for $i\ge1$ because otherwise chain $B$ would have been overtaken by chain $B'$ earlier. Obviously, then $coverage\Delta(B'_0) > coverage\Delta(B_0)$.

We can look at $M$ as a function $M(B'_0, \dots B'_{N-1})$ of the vector $B'$. We want to assess under what values of $B'$ the function $M$ reaches the lower bound, which would be the most comfortable situation for the attacker \textit{Eve}. This leads to a linear programming problem with the constraints given in the previous paragraph and the objective function $M(B'_0, \dots B'_{N-1})$, which should be minimized. We \textbf{conjecture} here that $M(B'_0, \dots B'_{N-1})$ reaches the minimum when $coverage\Delta(B'_i) = coverage\Delta(B_i)$ for $i=1\dots N-1$.

Based on the above, it means that the best scenario for the attacker is when all "malicious" tokens are concentrated in branch $B'_0$, however, she will need to secure enough amounts of malicious capital to satisfy $coverage\Delta(B'_i) = coverage\Delta(B_i)$. Thus, the lower bound $\mu$ of the amount of "malicious" tokens needed to overtake $B_0$ is given by
$$
\mu = \min_{0\le i\le N-1}\{2\cdot coverage\Delta(B_i) - totalSupply\} 
$$

Note that assessment of the lower bound of "malicious" capital needed for the attacker does not give the full picture, because it is reasonable to assume that tokens cannot easily "collude" when ownership of tokens is split over multiple (at least hundreds) sequencers and other token holders. This must be an assumption of the full model. Also, all the values of $coverage\Delta(B_i)$, as well as the "innocent" tokens $H$ that are isolated from the rest, must be assumed to be random variables. 

We are unable to build a complete model that also takes into account the randomness of the process; therefore, here we limit ourselves to the following \textbf{conjectures}:
\begin{itemize}
    \item In the optimal situation for the attacker, \textit{Eve} must ensure the amount of "malicious" capital close to the overtaking threshold \textbf{in every slot}. This is not difficult due to fragmentation of the token holdings; See also consideration in case of high concentration of the capital in \ref{sec:social_consensus}
    \item The probability $P$ of consolidating enough malicious capital in a slot is small. The probability of reaching that amount in the $N$ slots is $O(P^N)$, that is, it is decreasing exponentially with the length of the chain of branches which contain $T$. 
\end{itemize}

%============================================================
\subsection{Liveness}
\label{sec:liveness}

The confirmation rule introduced in the previous section requires at least $2\cdot \theta \cdot \text{totalSupply}$ of coverage actively contributing to the ledger coverage. If less than $\theta$ of the total on-ledger capital participates, \textit{Bob} may be waiting indefinitely for $\beta_T(B_0, \theta)$ to reach the desired value. He will perceive the situation as a \textbf{liveness problem}, meaning his transactions would cease to be confirmed.

Inflation is a stringent incentive for tokens to participate in the ledger coverage. It is feasible to expect high capital participation in the consensus, possibly approaching $totalSupply$.

\subsection{Balancing safety and liveness with $\theta$}
\label{sec:safety-and-liveness}

The subjective \textit{finality rule with parameter $\theta$} (see \ref{def:finality}) assumed by the user \textit{Bob} who issues the transaction $T$ means he is waiting for the \textbf{proof} that $T$ was included in the ledgers of other token holders, which together amounts at least $\theta$ part of the \textit{totalSupply}.

By assuming small values of $\theta$ (close to $1/2$) means \textit{Bob} faces significant risk, the transaction will be reverted\footnote{that would be equivalent of waiting only for 1 block in the PoW chain}.

Taking $\theta$ close to 1, \textit{Bob}, if lucky, can reach very high safety of his transaction because then $T$ could be reverted only by the consolidated effort of almost all token holders on the network, which is very unlikely.  

So, for safety, \textbf{the larger the parameter $\theta$, the safer the transaction}. 

On the other hand, large values of $\theta$ means \textit{Bob} will be waiting until  almost all token holders include $T$ in their ledgers. That may never happen and \textit{Bob} will never convince himself that his transaction $T$ is final.

Therefore, for liveness, \textbf{the smaller the parameter $\theta$, the easier the network is to perceive as live}.

We conclude that the \textbf{parameter $1/2<\theta<1$ balances the user's assumptions about safety and liveness}. The optimal value of $\theta$ must be somewhere in between the extremes, for example $2/3$ or $3/4$.

It is interesting to compare this finality rule with the \textit{longest chain rule} in PoW protocols. In the safety part, the rules are somehow similar and both assume an honest majority of 50% for the network to be secure.

However, the liveness aspect is not part of the longest chain rule, because there is no objective proof which part of the hashrate participated in finding the PoW nonce. 

\subsection{Network partitioning}
\label{sec:network_partitioning}

When players (peers) in the distributed ledger cannot communicate with the rest of the network, we have the situation of dividing (partitioning) the network into several parts. We want to analyze implications of partitioning for the security of the ledger in Proxima.

When a network is partitioned due to communication failures between miners (in \gls{pow}), validators (in \gls{pos}, \gls{bft}), or token holders (in \gls{cooperative consensus}), each network behaves differently. We will roughly compare how system behaves in each of the three principles described above. Our intention is to point to the fundamental differences between approaches with respective assumptions and to position Proxima with this regard. 

\subsubsection{BFT, PoS}
\label{sec:bft_pos}

The \gls{pos} \gls{bft} setup always assumes a bounded set of validators, a committee. A committee will continue operating as long as the largest partition of the network contains at least $\lfloor 2/3 \rfloor + 1$ of validators. If this condition is not met, the system stops producing blocks, leading to a liveness issue. However, the system remains secure in the sense that the transactions included in the ledger are considered valid and irreversible\footnote{this is highly simplified model, because many PoS systems employ intricate committee rotation mechanisms which makes liveness analysis not so straightforward}. This property is known as \textit{deterministic consensus}.

It is said, that \textbf{BFT-based systems prioritize safety over liveness}.

\subsubsection{PoW. Bitcoin and similar}
\label{sec:pow_btc}
In \gls{pow} systems such as Bitcoin, the response to network partitions differs significantly from \gls{bft} systems. If the miners are isolated into separate partitions (for example, for political reasons), each partition will continue to produce blocks at a reduced rate (due to a lower total hashrate in each partition) until the PoW difficulty adjusts accordingly. Consequently, each partition will maintain its own version (fork) of the ledger. This situation can persist unnoticed until the network reunites (which will mean that all but one fork will be \gls{orphaned} and \glspl{ledger state} reverted).

It is said, that \textbf{Nakamoto systems prioritize liveness over safety}. Users risk having their transactions included in a partition that might later be reverted. 

In the PoW systems, there is no objective way to assess which part of the global hashrate is contributing to the security of the user's partition. This property is an implication of the anonymity of the players in the consensus. 

It is important to note that Nakamoto (PoW) systems possess an automatic "self-healing" capability once communication barriers are resolved. Conversely, the process of "healing" in complex BFT/PoS systems may involve more intricate mechanisms, which require social consensus among the players.

\subsubsection{Safety and liveness in Proxima}
\label{sec:safety-proxima}
In \gls{cooperative consensus}, the behavior during network partitioning - specifically among token holders -- is akin to that observed in \gls{pow} systems. 

When isolated into partitions, each partition in Proxima will continue to follow the biggest ledger coverage rule, maintaining its own chain of branches at the pre-split pace. This results in each partition representing its own \textbf{fork} of the ledger state. 

Upon reconnecting the network partitions and resuming the transaction exchange, the nodes will synchronize their UTXO tangles. The \gls{biggest ledger coverage} rule dictates that sequencers in partitions with a smaller \gls{ledger coverage} must revert their chains and start anew from the \gls{baseline state} of the branch with the highest ledger coverage. Forks with lesser coverage will be \gls{orphaned}, illustrating a form of "self-healing."

Thus, cooperative consensus in Proxima exhibits behavior similar to PoW consensus systems, \textbf{prioritizing liveness over safety}, which is characteristic of Nakamoto consensus.

However, there are crucial safety-related distinctions between cooperative consensus and PoW. 

In PoW systems such as Bitcoin, users typically employ a "6 block rule" to ensure transaction safety, assuming the transaction is secure once buried under six blocks. This indirectly relies on the assumption that the user's chain is part of the partition with at least $51\%$ of the global hashrate, a presumption lacking objective verification.

In \gls{cooperative consensus}, token holders will adopt a similar acceptance rule: waiting until all branches with ledger coverage exceeding $51\%$  of the total token supply include their transaction $T$ (see Section \ref{sec:safety}.\nameref{sec:safety}). 

Unlike PoW, Proxima users know the partition's token weight, allowing them to withhold confirmation until a sufficient majority threshold is met. This means that if the partition is smaller than $51\%$ of the total supply, the user will not consider their transaction confirmed and will wait for the network partitions to reunite. 

Therefore, the \textbf{perception of liveness in Proxima is tightly linked to safety assumptions}. 

\begin{observation}
In \gls{cooperative consensus}, users consider the network non-operational until their transactions are deemed safe by the dominating amount of capital to approve it. \textbf{This is different from PoW}.
\end{observation}

Moreover, Proxima introduces considerations of malicious behavior among token holders. Security measures rely on economic incentives, where each token holder acts in their most profitable interest. This setup assumes legitimate behavior but requires precautions against economically motivated attacks, such as \textbf{long range attack} described in Section \ref{sec:safety}. 

To attack the network, the attacker would have to be a large token holder with significant funds in their possession. That significant stake implies considerable financial risk for the attacker itself, dissuading most attackers due to potential losses outweighing gains.

Analogous to why 2 or 3 dominant Bitcoin mining pools do not collude to alter ledger rules like halving, Proxima's ledger security hinges on significant token holder commitment, emphasizing the importance of "skin-in-the-game" among stakeholders.

In summary, Proxima's cooperative consensus model merges aspects of PoW robustness with mechanisms for managing safety and liveness during network partitions, incorporating economic incentives to maintain network integrity.

\subsection{Scalability of accounts and liveness}
\label{sec:scalability-liveness}

\subsubsection{Delegation}
\label{sec:scalability-delegation}

The cooperative consensus assumes that a significant majority of tokens must be participating in the consensus all the time. It effectively means that most of tokens must be moved constantly, in each slot.  

This fact presents a unique scalability challenge for the protocol, which is absent in other crypto ledgers. We envision at least millions of user accounts that should move their tokens constantly, so how exactly could we do that?

To address this challenge, we introduce the trustless ledger primitives of \textbf{frozen coverage} and \textbf{delegation}. The ledger primitives present special ledger validity constraints (\textit{covenants}), among many others, and are closely related to the concept of \glspl{chain}.

Here we will describe the delegation covenant of \textit{frozen coverage} and \textit{delegation} in general, without going into detail.

The token owner wraps their tokens in the chain output, called the \textbf{delegation output}, and locks it with the \textbf{delegation lock}. The latter allows the specified target sequencer to consume the delegation output and \textbf{freeze} it for a specified but bounded number of slots. For the frozen period of time, only the sequencer can consume the frozen output and only with the purpose of unfreezing it. The frozen funds in the delegation output do not move during the freeze period.

The sequencer is allowed to include the frozen amount in their ledger coverage for the freeze period. In addition, the frozen tokens generate inflation for the sequencer in each slot. 

To compensate the delegator for the loan capital, the sequencer must deposit an \textbf{inflation advance} to the delegation output upon freezing it. The protocol of how exactly inflation rewards are shared between the delegator and the sequencer is enforced according to the parameters set by the delegator and the sequencer, as two negotiating sides. In general, the whole process is market driven: for example, if the delegator considers the inflation margin required by the sequencer too high, she moves to the cheaper sequencer (or with better reputation score).

The guarantee that the sequencer cannot steal funds loaned to it is encoded in the delegation covenant (constraint script). After each freeze period, the ledger enforces \textbf{safe revocation window} during which only the delegator can consume the output. So, the delegator is guaranteed full control over her funds after a freeze period.

The maximum freeze period is set by the delegator and can be between 1 and 12 hours. The safe revocation window is around 10 min. 

The safe revocation window is just an enforced guarantee for the delegator so that she could always take her funds back. In an overwhelming majority of cases though, the sequencer will return funds immediately upon request from the delegator. Note that in theory the sequencer may ignore the unfreeze request (sequencer is a token holder, a centralized entity). That is expected to be a rare case because by misbehaving, the sequencer will gain small amount short-term, yet it will suffer reputation costs and lost income long-term.

With the inflation and the delegation mechanism, we achieve the following.
\begin{itemize}
    \item each token holder is incentivized to delegate their holdings to sequencers of their choice, while enjoying full \textbf{liquidity of the delegated funds}\footnote{similar to the \textbf{liquid staking} in PoS};
    \item sequencer is including coverage of the frozen tokens in each slot without moving the UTXO more often than once per 1 to 12 hours. That results in \textbf{high liveness} and \textbf{high scalability} (with respect to the number of accounts) of the network;  
\end{itemize}

\subsubsection{Compulsory freezing}
\label{sec:liveness-delegation}

Participation in the consensus is crucial for the safety and liveness of the system. For this reason, all participants in the Proxima ledger are incentivized to participate in the consensus by design.  

However, there is no guarantee that all token holders will follow rational behavior for many reasons: neglecting the dilution costs, lost private keys, etc. 

To address this problem, we introduce special ledger rules that enable and incentivize any sequencer to freeze any UTXO that has not been moved for more than a certain number of slots (say 24 or 48 hours). The \textbf{compulsory freezing} follows the usual delegation rules with the trusted revocation, a maximum of 12 hours of freezing, and the safe revocation window. This guarantees that the owner will never lose her funds, yet the funds will participate in the consensus. 

In this way, we ensure high liveness of the system and high participation of the capital in the safety of the ledger without compromising scalability.

%====================================================================================
\subsection{Ultimate security. Social agreement}
\label{sec:social_consensus}

Token holders exchange transactions, which is the only category of messages on the network. Sequencers are distinct from other token holders only in the specific way they contribute to the shared ledger.

Token holders who control the dominant majority of the on-ledger capital define the global consensus on the ledger state. The rest of the network follows the ledger version with the largest coverage, which is secured by the majority.

A set of token holders controlling the dominant majority of the total supply can decide on any option allowed by the ledger validity constraints: censoring transactions, creating forks, running parallel forks, abandoning one fork, continuing with another, and so on. As in all distributed ledgers, the protocol cannot prevent this, usually undesirable, possibility. However, in the cooperative consensus of Proxima, forking and stitching forks together is a routine, low-cost process by design.

The \textit{security aspect} of a distributed ledger is closely related to \textit{decentralization}, which in turn is closely associated with \textit{permissionless writing to the ledger} (\textit{open participation}), although these concepts are not equivalent.

\begin{observation}
Over time, open participation usually leads to high concentration of influence on the network, such as hashrate in PoW networks, which may result in less decentralization rather than more.
\end{observation}

In BFT PoS networks, joining and leaving the committee of validators is not as straightforward, which paradoxically can lead to more decentralized influence compared to completely permissionless PoW networks (although this also depends on other factors, such as how exactly decentralization of the committee is ensured). There are no ultimate decentralization metrics in a BFT/PoS system.

The observation above is also consistent with the \textbf{demonstration of the social agreement} among the dominant Bitcoin miners, who understand that the vast majority of Bitcoin's real value stems from trust in the network and its expected future. 2-3 largest block proposers (miners) could easily fork the network, for example, with the purpose of changing halving rules. However, they are aware that undermining the trust publicly by compromising decentralization through a protocol fork supported by the majority of hashrate would have unknown consequences, likely resulting in irreversible damage to the Bitcoin narrative and substantial losses for everyone invested in it. Therefore, \textbf{the fundamental source of Bitcoin's myth (and, ultimately, its market value) is the strong social agreement among dominant miners}. The ledger consensus protocol serves as a tool for coordinating that real-world (social) consensus.

This leads us to another universal thesis: 
\begin{observation}
The security of any distributed ledger is proportional to the vested interest of (quantifiable) amount of skin-in-the-game  committed by a limited set of dominant stakeholders.    
\end{observation}

The following statement can be useful as an universal assumption about the ultimate cause of security in the distributed ledgers\footnote{Despite of the imprecise definition and speculative nature of the very idea of the \textit{social consensus}}:
\begin{assumption}
The security of the distributed ledger ultimately depends on the \textbf{social agreement} among entities with the dominant skin in the game.     
\end{assumption}

The \gls{cooperative consensus} will very likely follow universal patterns and converge to a high capital concentration, similar to other fully open participation networks.

As stated earlier, the Proxima design assumes that capital concentration is inevitable and that the network will ultimately be run by a limited number (probably hundreds, even though the protocol does not set any limits) of sequencers, supported by a substantial amount of capital delegated to sequencers by other token holders. In fact, the amount of delegated capital is likely to dominate the capital owned by the sequencers\footnote{we can see strong analogies with how banking sector operates, which normally employs much more leveraged capital than equity in their balance sheets}. 

So, the fate of the distributed ledger will depend on the social consensus among these dominant players, such as major crypto exchanges, established L2 chains, and other capital custodians that attract other capital through delegation by non-sequencing token holders.

The bootstrap process of such a network begins with the distribution of the genesis supply to a selected group of initial token holders, typically around half a dozen to a dozen. These initial token holders will launch sequencers following the initial coin offering (ICO). The further decentralization of the network will then be driven by the dynamics of the market.

%==========================================================================
\section{Denial-of-service prevention} 
\label{sec:anti-spam}
The \textit{denial of service} attacks aim to overwhelm the network and disrupt its operation by sending it large amounts of messages. In Proxima those messages are transactions. We identify several different targets for such attacks and respective countermeasures.

In general, we rely on two main principles for the DoS-prevention: \textbf{Sybil resistance} and \textbf{limit on transaction-per-second rate per user}. 

The \textit{Sybil resistance} means, that in related attack vectors, the attacker will need to split its holdings among many different fake token holders (addresses) and that will incur cost of storage deposit (see below).

%==========================================================================
\subsection{State bloat}

The ledger \textit{state bloat} refers to the act of issuing many small-amount outputs, commonly known as \textbf{dust}. The large amount of dust is highly undesirable because it inflates the size of \gls{ledger state} with a large number of useless \glspl{UTXO} that never move, because the cost of consumption exceeds the value of locked tokens. It is also a potential attack vector.  

Proxima prevents state bloat by enforcing a minimal number of tokens on each UTXO, based on the size of it in bytes. This is known as a \textbf{storage deposit}. For a token holder, the storage deposit becomes a minimal amount of its holdings. The storage deposit is "refunded" when multiple outputs are compacted into one output. 

The minimal storage deposit may lead to suboptimal consequences in the context of \glspl{tag-along fee}, which may be very small or even zero. Using an ordinary \textit{chain-locked} output as a tag-along output, a portion of the output is essentially paid as a fee to the sequencer. This can potentially create problems, such as

\begin{itemize}
    \item The fee cannot be less than the storage deposit, which may result in an excessively high fee;
    \item If, for some reason, the target sequencer never consumes \gls{tag-along output}, but the transaction is recorded in the ledger for other reasons, the storage deposit may be lost for the sender. 
\end{itemize}

To address the problem, a special type of lock constraint script called the \textit{tag-along lock script} is introduced. It possesses the following properties:
\begin{itemize}
    \item There is no requirement for the minimum amount on the tag-along output; it can also be a zero-amount.
    \item The target chain can consume a tag-along output for the first 12 slots (approximately 2 min).
    \item If it is not consumed within 12 slots, the sender can unlock it during the next 100 slots. This allows the sender to recover any unconsumed tag-along fee.
    \item If the output is still not consumed after 100 slots, it becomes consumable by any transaction. This enables a permissionless "vacuum cleaning" process to remove any remaining dust from the ledger.
\end{itemize}

%==========================================================================
\subsection{Spamming the ledger}

By sending tokens in a consume/produce loop, one can issue an overwhelming number of chained transactions that would all be included in the ledger on the UTXO tangle yet without spamming the ledger state. That could negatively impact node performance.

To prevent this attack vector, we introduce the \textbf{transaction pace} constraint for transaction timestamps. The \textit{transaction pace} is a ledger constant that defines the minimum number of ticks between consumed and consumed transactions, as well as between endorsing and endorsed transactions. This limits the number of chained transactions that can fit in one slot. It is reasonable to have different paces for sequencer milestones and user transactions: smaller for sequencers and more significant for users\footnote{in the testnet implementation sequencer milestone pace is 1 tick, non-sequencer transaction pace is 25 ticks (~10 transactions per slot)}.

%==========================================================================
\subsection{Spamming the UTXO tangle}

Spamming with conflicts cannot be prevented by ledger constraints, as an attacker could produce potentially unlimited amounts of double-spends from one output and post them to the UTXO tangle. In this way, \gls{UTXO tangle} will have to insert an unbounded number of vertices. It is an attach vector.

Our approach to prevent spamming the UTXO tangle is to implement rate limits per user (per \gls{token holder}). This is possible because each transaction (which is also the message between nodes) is signed and identified by the sender's public key.

The nodes are responsible for enforcing the rate limits. If a node receives several transactions from the same sender in a real-time period equal to the transaction pace according to its local clock, subsequent transactions from the user are delayed as a form of punishment. The strategy must be applied subtly; for example, transactions should not be punished with delay if they were pulled by the node itself.

Repeated chain transfers with the small interval will cause them to be \gls{orphaned}. If, due to asynchronicity of the network, some transaction will be removed from some nodes and will remain on other nodes, the consistency of the tangle will be restored by the routine \textit{solidification mechanism}. This is possible because \textit{ cooperative consensus}, being a \textit{probabilistic consensus} with non-deterministic finality, does not rely on the concepts of transaction being confirmed or rejected. 

This punishing strategy would limit transaction rates per \gls{token holder} to, for example, one transaction per second or even less, which is reasonable for non-sequencer transactions. The natural Sybil-protection of token holders and minimum storage deposit means that every attempt to bypass the rate limiter by creating many accounts will have the cost for the attacking entity.

Determining the optimal constants for the minimum storage deposit and the rate limits is subject to simulation and modeling. \\ \\ \\

Vilnius, 2024, 2025

\clearpage
\printglossaries   % [title=Used Terms, toctitle=List of Used Terms]

\end{document}